\documentclass[twocolumn,showpacs,preprintnumbers,amsmath,amssymb]{revtex4}

\usepackage{graphicx}
\usepackage{hhline}
\usepackage{dcolumn}% Align table columns on decimal point
\usepackage{bm}% bold math

\usepackage[dvips]{color}

\begin{document}

\title{Numerical study of a three-state
host-parasite system on the square lattice}

\author{Takehisa Hasegawa}
\email{hasegawa@stat.t.u-tokyo.ac.jp} 
\affiliation{
Department of Mathematical Informatics,
The University of Tokyo,
7-3-1 Hongo, Bunkyo, Tokyo 113-8656, Japan
}
\author{Norio Konno}
\email{konno@ynu.ac.jp} 
\affiliation{%
Faculty of Engineering, Yokohama National University, 79-5 Tokiwadai, Hodogaya-ku, Yokohama,
%240-8501
 Japan
}%
\author{Naoki Masuda}
\email{masuda@mist.i.u-tokyo.ac.jp} 
\affiliation{
Department of Mathematical Informatics,
The University of Tokyo,
7-3-1 Hongo, Bunkyo, Tokyo 113-8656, Japan
}
\affiliation{PRESTO, Japan Science and Technology Agency,
4-1-8 Honcho, Kawaguchi, Saitama 332-0012, Japan}
%\date{\today}

\begin{abstract}
We numerically study the phase diagram of a three-state 
host-parasite model on the square lattice motivated by population biology.
The model is an extension of the contact process, and the three states correspond to
an empty site, a host, and a parasite. We determine 
the phase diagram of the model by scaling analysis.
In agreement with previous results, 
three phases are identified: the phase in which both hosts and parasites
are extinct ($S_0$), the phase in which hosts survive but parasites
are extinct ($S_{01}$), and the phase in which both hosts and parasites survive
($S_{012}$). 
We argue that both the $S_{0}$--$S_{01}$ and $S_{01}$--$S_{012}$ boundaries belong to the directed percolation class.
In this model, it has been suggested that
an excessively large reproduction rate of parasites 
paradoxically extinguishes hosts and parasites and results in $S_0$.
We show that this paradoxical extinction is a finite size effect;
the corresponding parameter region is likely to 
disappear in the limit of infinite system size.
\end{abstract}
\pacs{87.23.Cc, 05.50.+q, 64.60.an}% PACS, the Physics and Astronomy
\maketitle

\section{introduction}

In research fields ranging from ecology and epidemiology to sociology, 
it is important to clarify 
the effect of the interactions among species or phenotypes on the entire system.
Stochastic interacting particle systems, in which
each site on a graph takes either of the possible states and is flipped 
according to the states of other sites, 
are a useful tool for this purpose.
A paradigmatic interacting particle system that describes disease spreading
is the contact process (CP; also termed the
susceptible-infected-susceptible model)
\cite{marro1999nonequilibrium,liggett-interacting,durrett1994stochastic}.

Various interacting particle systems
in complex networks have been investigated recently \cite{newman2003structure,barrat2008dynamical}.
Nevertheless, in an ecological context,
organisms of different scales can be considered to live in
a two-dimensional space, often with a small interaction range. 
Therefore, it is instructive to
study models that are more complex than the CP on the Euclidean lattice
 \cite{dieckmann2000geometry,anderson1991infectious,durrett1994stochastic}.
A simple extension of the CP in this direction is a three-state
spatial host-parasite (HP) model that deals with an ecosystem comprising
soil (empty sites), host species living on
soil, and pathogen species (parasites) living on hosts.  Phase
transitions and oscillations in similar models have been
studied from the perspective of statistical physics 
\cite{antal2001critical,antal2001phase,de2004probabilistic,de2006self,peltomaki2005host}.

Sat{\=o} et
al. \cite{sato1994pathogen} analyzed the HP model on a square lattice. 
They showed by means of the improved-pair approximation (i-PA) and numerical simulations
that
a very high reproduction rate of parasites 
results in the extinction of both hosts and parasites. 
This phenomenon is called parasite-driven extinction 
\cite{sato1994pathogen,haraguchi2000evolution}.
An intuitive explanation for this paradoxical behavior is that
parasites replace hosts so quickly that hosts get extinct, which
eventually results in the extinction of parasites. 
A similar paradoxical behavior, 
i.e., a decrease in the number of a species
caused by an increase in its fertility, 
is observed in other models,
where a sort of rock-scissors-paper competition is prevalent among three species 
\cite{boots2000evolutionary,boots2002parasite,boots2003parasite,tainaka1993paradoxical,tainaka1995indirect,frean2001rock,ryder2007host,lipowski1999oscillatory,kowalik2002oscillations}.
However, the current understanding of the
phase diagram of the HP model is not comprehensive, 
because parasite-driven extinction cannot be predicted by 
mean field approximation and pair approximation (PA)
\cite{ohtsuka2006phase,sato1995successional}.

\begin{figure}
\begin{center}
\includegraphics[width=8cm]{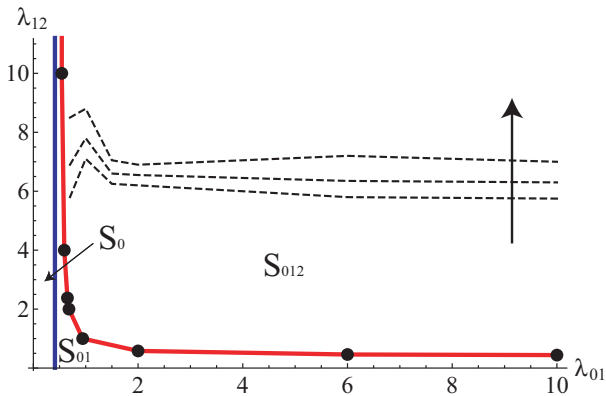}
\end{center}
\caption{
Phase diagram of the HP model.
The solid blue line
represents the boundary between $S_0$ and $S_{01}$
 (i.e., $\lambda_{01}=\lambda_{01}^{\rm c}$).
The solid red line represents the boundary between $S_{01}$
and $S_{012}$ and is drawn on the basis of the data shown in
Table~\ref{table1}.
Above the dashed lines, which correspond to
$L=300,500$, and 1000 from the bottom to the top,
the parasite-driven extinction occurs frequently.
}
\label{phasediagram}
\end{figure}

In this paper, we numerically investigate the phase diagram of the HP model 
on the square lattice. In particular, we use large lattices and investigate
the effect of the system size on parasite-driven extinction. 
%by means of finite size scaling.
The obtained phase diagram is shown in Fig.~\ref{phasediagram}.
We argue that two transition
boundaries (solid lines in Fig.~\ref{phasediagram})
belong to the directed percolation (DP) universality class. 
Another transition boundary (dotted lines) is not characterized by
the DP universality class, and its location depends on the system size.
We claim that the parasite-driven extinction phase is a finite
size effect and that the phase diagram is qualitatively
the same as that obtained by the PA rather than that obtained
by the i-PA. 

%%%%%%%%%%%%%%%%%%%%%%%%%%%%%%%%%%%%%%%%%%%%%%%%%%%%%%%%%%%%%%%%%%%%%%%%%%%%%%%%%%%%%%%%%%%%%%%%%%%%

%%%%%%%%%%%%%%%%%%%%%%%%%%%%%%%%%%%%%%%%%%%%%%%%%%%%%%%%%%%%%%%%%%%%%%%%%%%%%%%%%%%%%%%%%%%%%%%%%%%%

\section{model}\label{sec:model}

\begin{figure}
\begin{center}
\includegraphics[width=6cm]{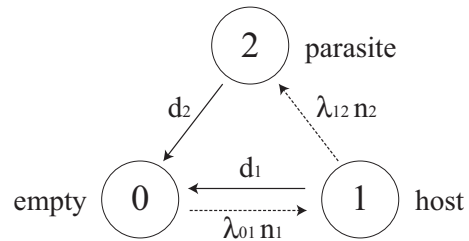}
\end{center}
\caption{ 
Transition rules of the HP model.
Solid and dashed lines represent deaths and births, respectively.
The values indicate the transmission rates, and
$n_i$ denotes the number of neighbors of a site in state $i$.}
\label{rule}
\end{figure}

The HP model on the square lattice $\mathbb{Z}^2$ is defined as a
continuous-time Markov process with state space $\{ 0,1,2
\}^{\mathbb{Z}^2}$ \cite{sato1994pathogen,sato1995successional,haraguchi2000evolution,ohtsuka2006phase,andjel1996complete}.
Each site takes one of the three states $0, 1,$ and $2$, which
represent an empty site, a host, and a parasite, respectively.
The rules for the state transition are depicted in Fig.~\ref{rule}.
A host and a parasite die at rates $d_1$ and $d_2$, respectively.  
For simplicity, we set $d_1=d_2=1$.  
The occurrence of death at any site is independent of the states of the neighboring sites. 
In contrast, reproduction of hosts and parasites depends on the states of the neighbors. 
A host emerges at an empty site $i$ at rate $\lambda_{01}n_1(i)$, 
where $n_1(i)$ is the number of hosts in the neighborhood of site $i$. 
A host at site $i$ turns into a parasite at rate $\lambda_{12}n_2(i)$, 
where $n_2(i)$ denotes the number of parasites in the neighborhood of site $i$. 
We vary the values of $\lambda_{01}$ and $\lambda_{12}$ in the numerical simulations.
Because parasites feed on hosts,
the HP model allows the following three phases in the stationary state:
\begin{description}
\item[(i)] phase $S_0$, in which
hosts and parasites are extinct,
\item[(ii)]
phase $S_{01}$, in which hosts survive and parasites are extinct, and
\item[(iii)] phase $S_{012}$, in which both hosts and parasites survive.
\end{description}

%%%%%%%%%%%%%%%%%%%%%%%%%%%%%%%%%%%%%%%%%%%%%%%%%%%%%%%%%%%%%%%%%%%%%%%%%%%%%%%%%%%%%%%%%%%%%%%%%%%%%%%%

%%%%%%%%%%%%%%%%%%%%%%%%%%%%%%%%%%%%%%%%%%%%%%%%%%%%%%%%%%%%%%%%%%%%%%%%%%%%%%%%%%%%%%%%%%%%%

The HP model with $\lambda_{12}=0$ is equivalent to the CP.
In the CP, each site takes either state 0 or 1, 
and a death event ($1 \to 0$) and a reproduction event ($0 \to 1$) at site $i$ occur at rate $d_1=1$ 
and $\lambda_{01}n_1(i)$, respectively.
In the case of the CP on the square lattice, $S_0$ and $S_{01}$ are realized
when $\lambda_{01}$ is respectively smaller and larger than 
$\lambda_{01}^{\rm c}\approx 0.4122$ \cite{marro1999nonequilibrium}.

The phase diagram of the HP model on the square lattice
has been examined using the mean field approximation \cite{sato1994pathogen}; 
the PA, which accounts for pairwise state correlation \cite{ohtsuka2006phase}; 
and the i-PA, which calibrates the PA to account for
the aggregation of the same species in the space
 \cite{sato1994pathogen,haraguchi2000evolution}.
All of the three approximations
predict the existence of the three phases of the model.
In the mean field approximation and
the PA, the system is in $S_0$
if $\lambda_{01}$ is less than a critical value that is independent of
$\lambda_{12}$.
Otherwise, the system is in $S_{01}$ ($S_{012}$) when the value of 
$\lambda_{01}$ and $\lambda_{12}$ is sufficiently small (large). 
In the mean field approximation and the PA,
the boundary between $S_{01}$ and $S_{012}$ is given by
$\lambda_{12}=\lambda_{01}/(4 \lambda_{01}-1)$ and
$\lambda_{12}=(12 (\lambda_{01})^2+4\lambda_{01})/(36 (\lambda_{01})^2-4 \lambda_{01}-3)$, respectively \cite{ohtsuka2006phase}. 
In particular, only $S_0$ and $S_{012}$ exists when
$\lambda_{12} \to \infty$ in the mean field approximation.
In the PA, when $\lambda_{12} \to \infty$,
$S_0$, $S_{01}$, and $S_{012}$ appear in this order in the PA
as $\lambda_{01}$ increases.
The phase diagram obtained from the i-PA is qualitatively distinct
from those obtained from the mean field approximation and the PA.
When $\lambda_{12}$ is large,
the i-PA predicts $S_0$ regardless of the value of $\lambda_{01}$.
This result corresponds to the numerical observation that 
a large reproduction rate of parasites induces extinction of hosts and parasites \cite{sato1994pathogen, haraguchi2000evolution}.
We call this phenomenon the parasite-driven extinction.
The mean field approximation and the PA do
not predict the existence of the parasite-driven extinction.

\section{DP transition on the $S_{01}$--$S_{012}$ boundary for small $\lambda_{12}$ \label{secDP}}

In this section, we numerically examine 
the boundary between $S_{01}$ and $S_{012}$
for small values of $\lambda_{12}$ (the red
solid line in Fig.~\ref{phasediagram}). 
We carry out Monte Carlo simulations for the HP model 
on the square lattice with $N=L \times L$ sites, where $L=300$.
Periodic boundary conditions are assumed.
We run 500 realizations for fixed $\lambda_{01}$ and $\lambda_{12}$.
At the beginning of
each realization, each site independently takes 
state 0, 1, or 2 with equal probability.
We adopt an event-driven update algorithm
in which we select one out of all the possible events to occur 
with the appropriate probability for each time step.
Then, we increment the time by an appropriate amount.

\begin{figure}
\begin{center}
\includegraphics[width=7cm]{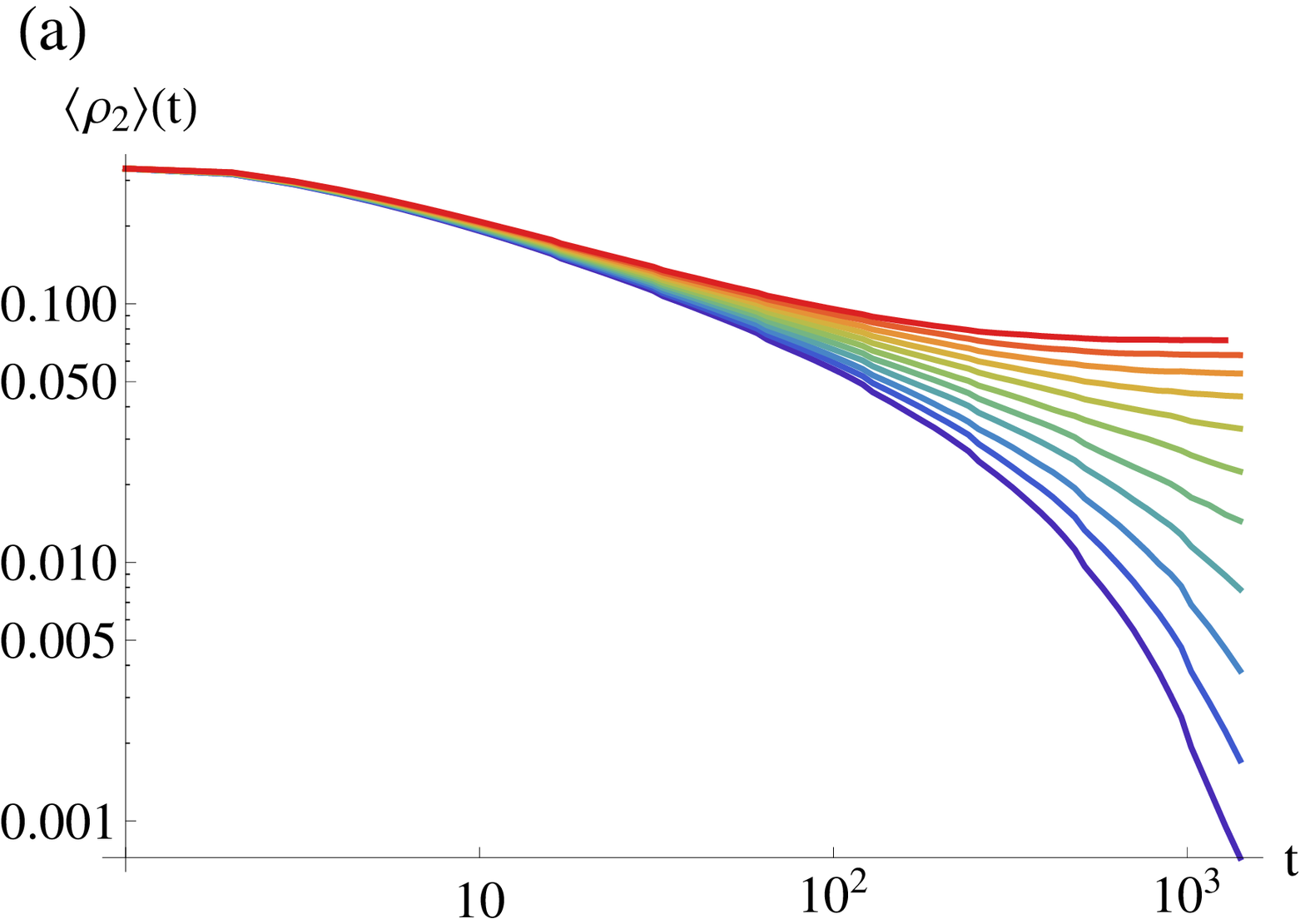}
\vspace{0.3cm}
\includegraphics[width=7cm]{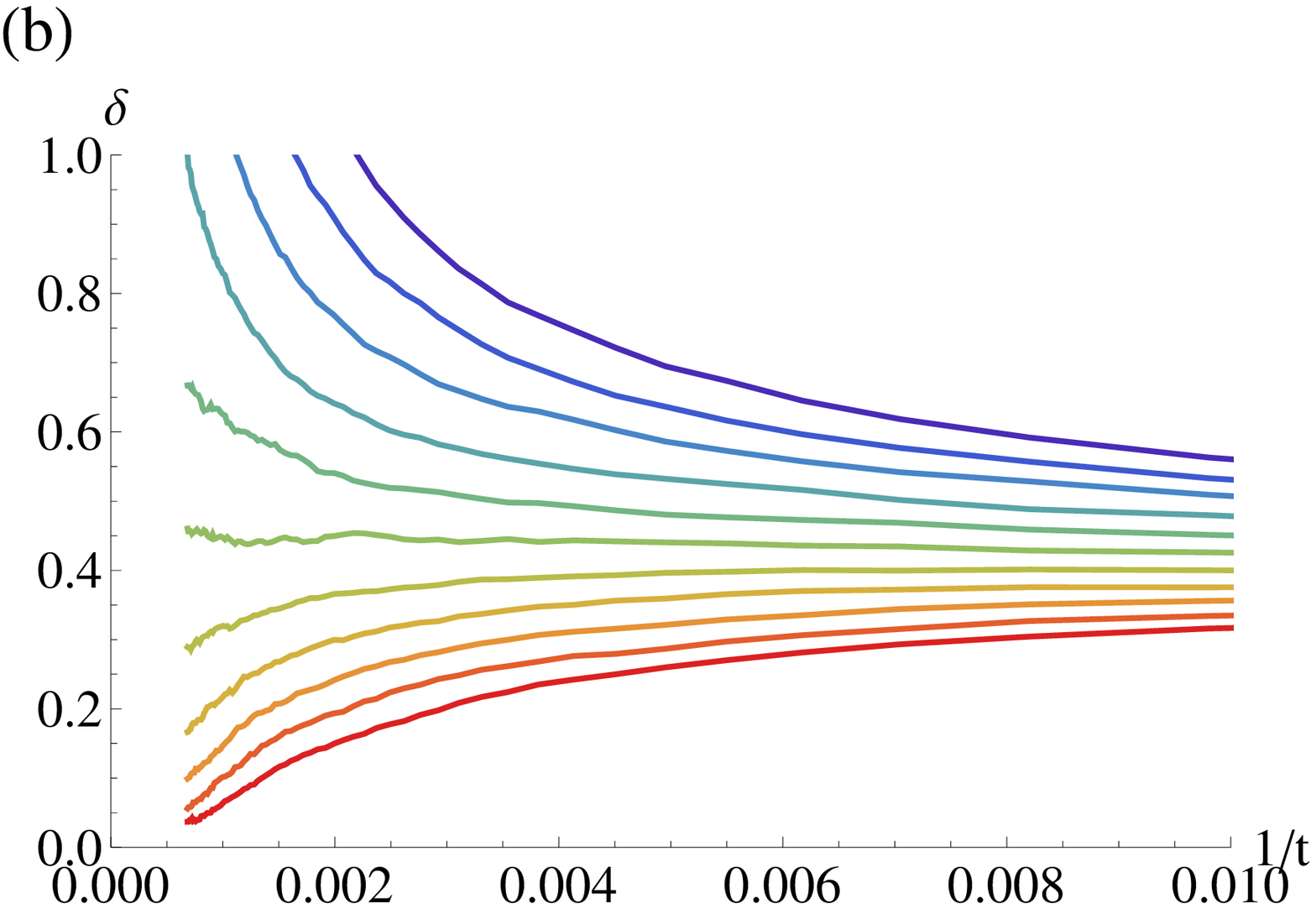}
\vspace{0.3cm}
\includegraphics[width=7cm]{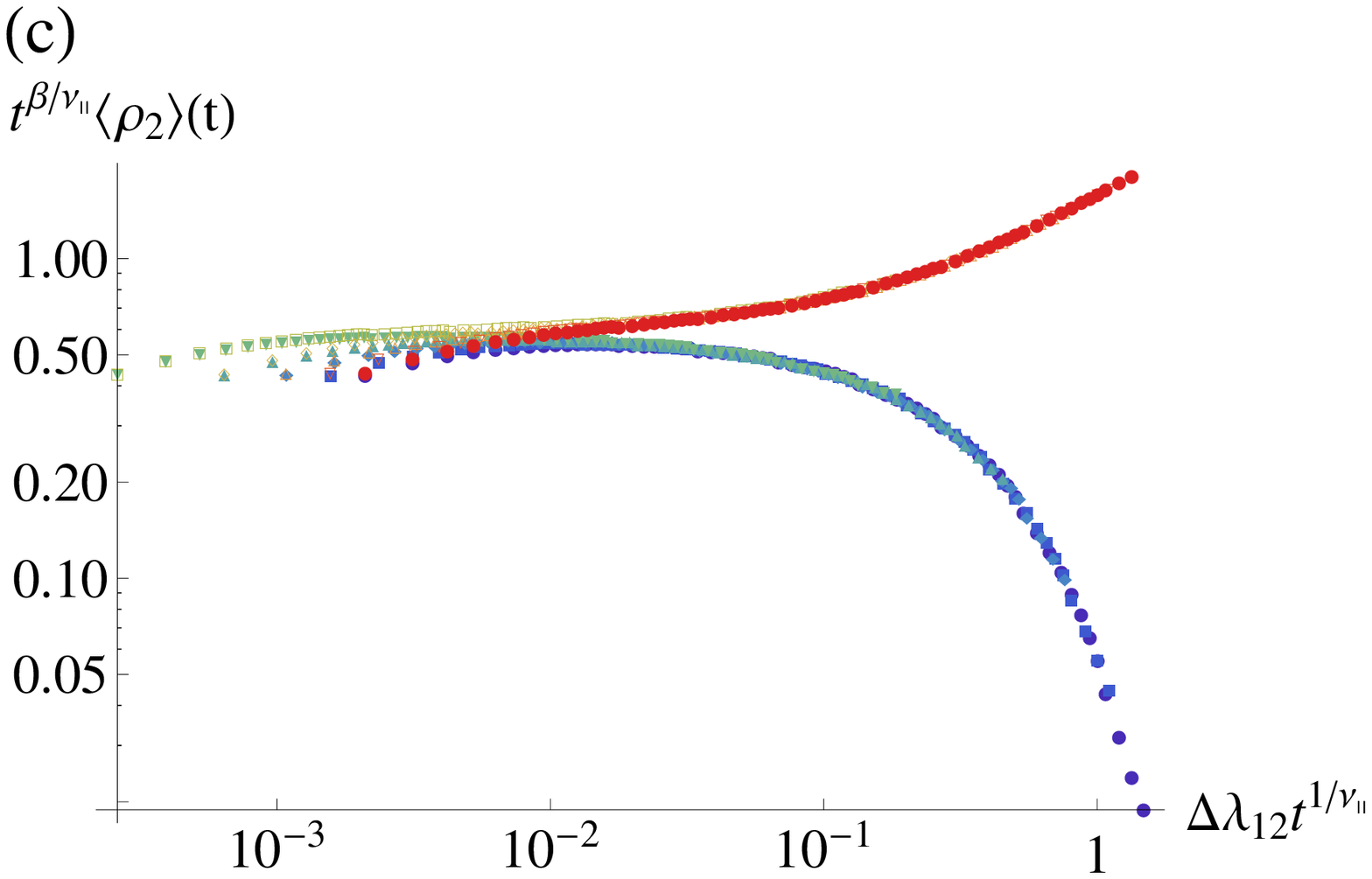}
\end{center}
\caption{DP phase transition at
$\lambda_{01}\to\infty$ and $\lambda_{12}\approx \lambda_{12}^{\rm c}$.
(a) Time courses of $\langle \rho_2 \rangle (t)$ and 
(b) local slope $\delta$ of $\langle \rho_2 \rangle (t)$.
The different lines from the top to the bottom correspond
to $\lambda_{12}=0.4079, 0.4089, \ldots$, and 0.4179.
(c) Dynamic scaling (Eq.~\eqref{scale1}) for the data shown in (b).
}
\label{p10infty}
\end{figure}

First, we focus on the limit $\lambda_{01}\to\infty$, 
where an empty site adjacent to a host 
is instantaneously replaced by the host.
A cluster of empty sites survives only when they are surrounded by
a shell of parasites.
When $\lambda_{12}$ is small, 
parasites rarely form such a shell.
Then, 
%there are few empty sites, and
the HP model behaves like the CP, where empty sites and hosts in the HP model
collectively correspond to the susceptible sites (i.e., state 0) in the CP.
Because many spatial stochastic processes including the CP 
undergo a DP-type phase transition \cite{hinrichsen2006non,hinrichsen2000non,Odor2004universality,marro1999nonequilibrium},
we expect that the HP model also undergoes a DP-type transition
from $S_{01}$ to $S_{012}$ as $\lambda_{12}$ is increased to cross
$\approx \lambda_{01}^{\rm c} \approx 0.4122$.
The time courses of the mean density of parasites $\langle \rho_2
\rangle (t)$ are
shown in Fig.~\ref{p10infty}(a) for various values of $\lambda_{12}$,
where $\langle \cdot \rangle$ denotes the average over
all the realizations.
At $\lambda_{12} =\lambda_{12}^{\rm c} \approx 0.4129$, 
we obtain
\begin{equation}
\langle \rho_2 \rangle (t) \propto t^{-\delta}.
\end{equation}
From Fig.~\ref{p10infty}(b), which shows the plotting of the local slopes of
$\langle \rho_2 \rangle(t)$, we obtain
$\delta \approx \log \langle \rho_2 \rangle(t)/ \log t
\approx 0.451$, a value indicative of the DP universality class
\cite{marro1999nonequilibrium}. 
%
%\cite{voigt1997epidemic}.
%
We also derive $\delta$ via 
dynamic scaling \cite{hinrichsen2000non,hinrichsen2006non}, i.e.,
by fitting the following scaling form:
\begin{equation}
\langle \rho_2 \rangle (t) \approx t^{-\beta/\nu_{||}} \tilde{\rho_2}
\left(\Delta \lambda_{12} 
t^{1/\nu_{||}}, \frac{t^{d/z}}{N}\right), \label{scale1}
\end{equation}
where
\begin{equation}
%\Delta_{12} = \lambda_{12}-\lambda_{12}^{\rm CP}.
\Delta \lambda_{12} = \lambda_{12}-\lambda_{12}^{\rm c}.
\end{equation}
The critical exponent $\delta$ is given by $\delta=\beta/\nu_{||}$. 
The results of the dynamic scaling 
with the known critical exponents for the 
(2+1)-dimensional DP universality class $\beta \approx 0.583$ 
and $\nu_{||} \approx 1.295$ 
\cite{marro1999nonequilibrium} %\cite{grassberger1996self} 
are shown in Fig.~\ref{p10infty}(c).
The data for different values of $\lambda_{12}$ collapse onto
a single curve separately for subthreshold and suprathreshold
values of $\lambda_{12}$. 
This result also supports that the transition belongs to the DP universality class.
%

%%%%%%%%%%%%%%%%%%%%%%%%%%%%%%%%%%%%%%%%%%%%%%%%%%%%%%%%%%%%%%%%%%%%%%%%%%%%%%%%%%%%%%%%%%%%%%%%%%%%

\begin{table*}
\caption{Several points on the $S_{01}$--$S_{012}$ boundary.
}
\label{table1}
\begin{tabular}{l|lllllllllllllll}
\hline
$\lambda_{01}$  &\,\, 0.509 &\,\,  0.543 &\,\,  0.591 &\,\, 0.651 &\,\, 0.680 &\,\, 0.942  &\,\, 2.000 &\,\, 6.000 &\,\, 10.000 &\,\, 15.000 &\,\, 20.000 &\,\, $\infty$ \\
\hline
$\lambda_{12}$ &\,\, $\infty$  &\,\,  10.000  &\,\,  4.000 &\,\, 2.378 &\,\, 2.000 &\,\, 1.000  &\,\, 0.581 &\,\, 0.459 &\,\, 0.440 &\,\, 0.430 &\,\, 0.426 &\,\, 0.4129 \\
\hline
\end{tabular}
\end{table*}

%%%%%%%%%%%%%%%%%%%%%%%%%%%%%%%%%%%%%%%%%%%%%%%%%%%%%%%%%%%%%%%%%%%%%%%%%%%%%%%%%%%%%%%%%%%%%%%%%%%%

If $\lambda_{01}$ is finite and sufficiently large,
we can numerically obtain the transition points and the critical exponents
in the same manner.
On the critical line, 
$\langle \rho_2 \rangle (t)$ shows a power law decay with $t$,
as shown in
Fig.~\ref{p10finite}(a).
When $\lambda_{01}\gtrsim 0.68$,
the dynamic scaling yields
the DP critical exponents at each examined transition point.
The locations of several points on the $S_{01}$--$S_{012}$ boundary
are shown in Fig.~\ref{phasediagram} and
Table~\ref{table1}.

We postpone the analysis of the case $\lambda_{01} \lesssim 0.68$ to Sec.~\ref{sec:6}.

\section{Dependence of boundary between $S_{012}$ and 
the parasite-driven extinction region on $\lambda_{12}$
\label{sec:4}}

\begin{figure}
\begin{center}
\includegraphics[width=7cm]{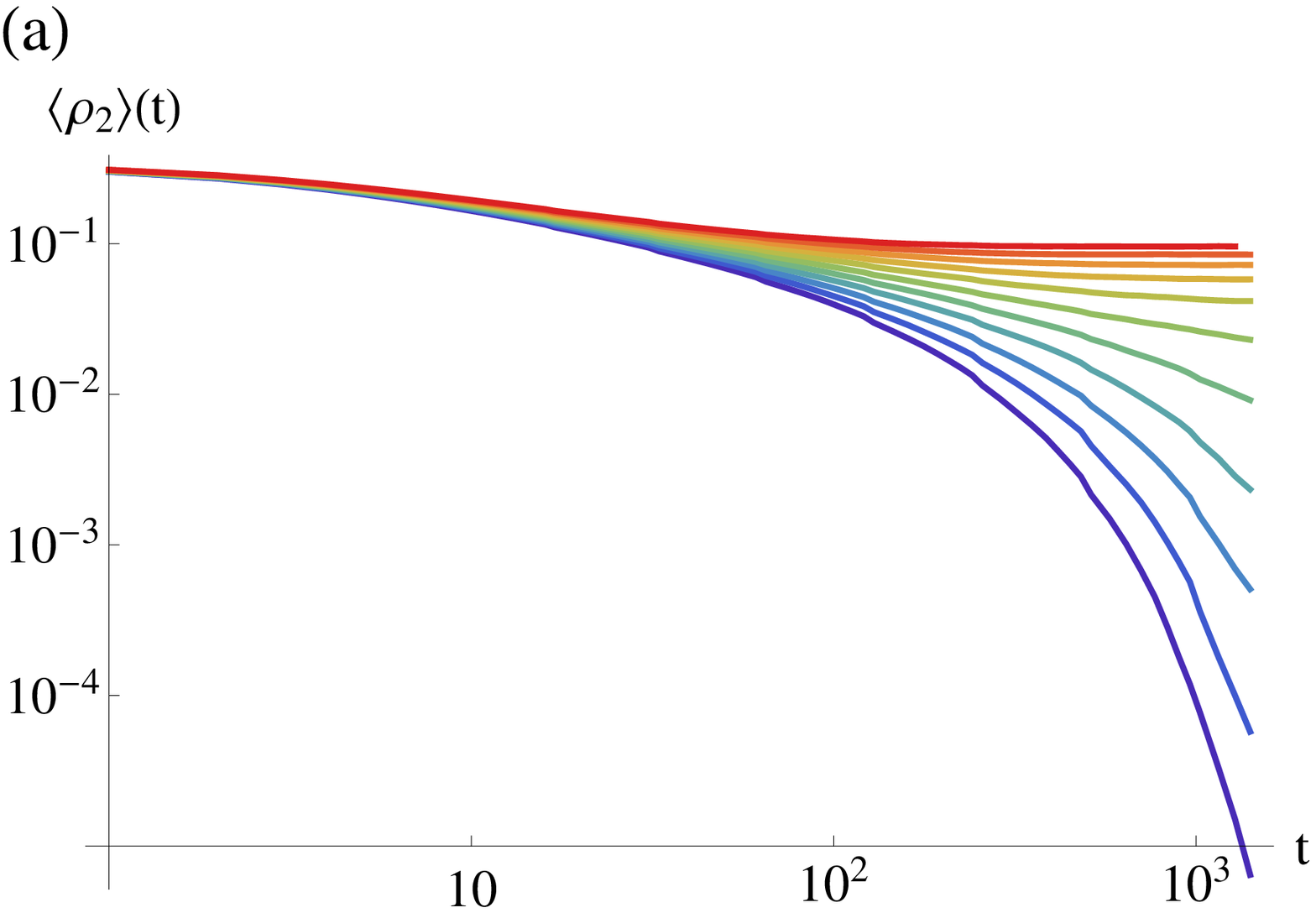}
\includegraphics[width=7cm]{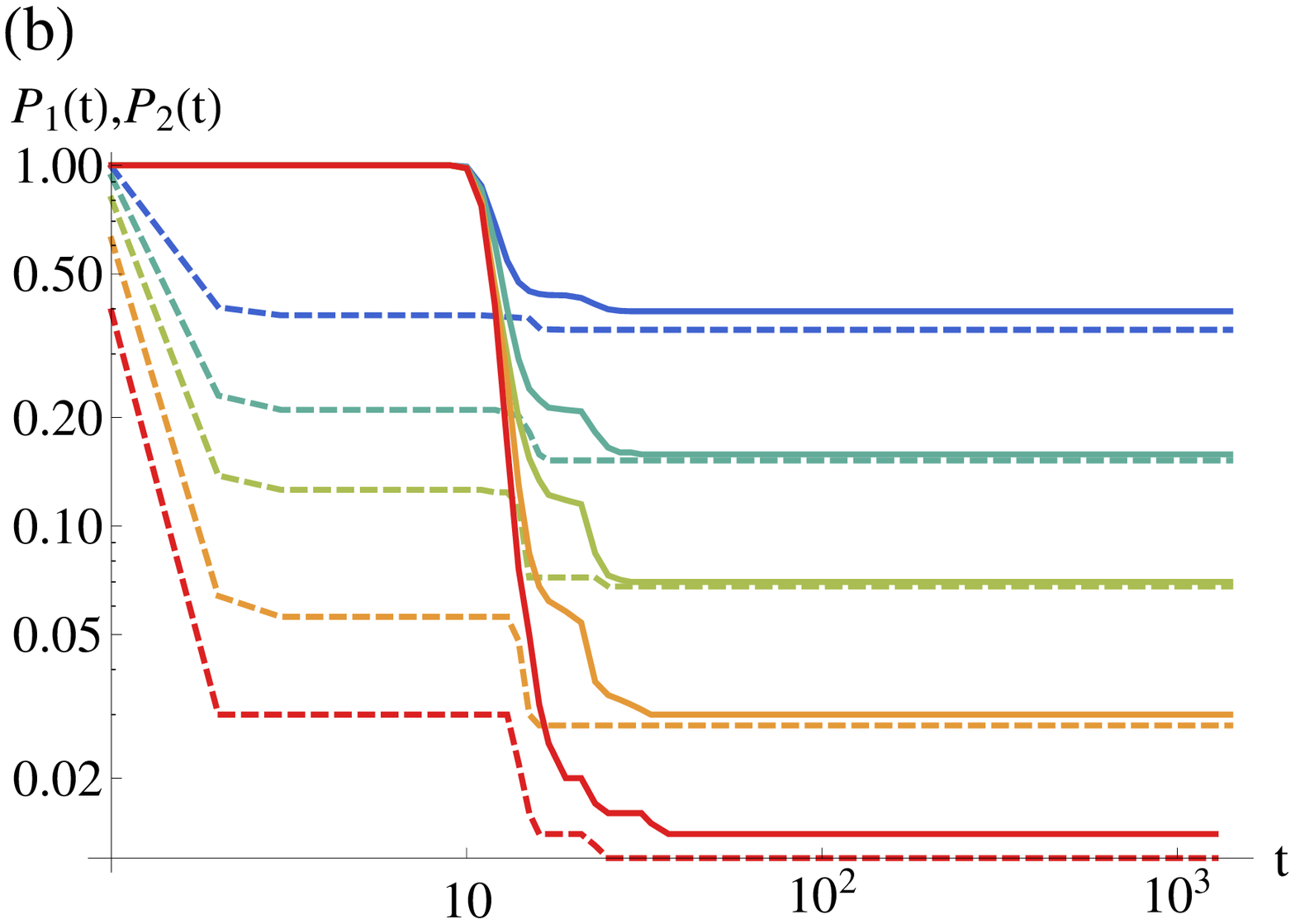}
\end{center}
\caption{
(a) Time courses of $\langle \rho_2 \rangle (t)$ 
with $\lambda_{01}=10$.
The lines from the top to the bottom correspond to
$\lambda_{12}=0.430, 0.432, \ldots$, and 0.450.
(b) Surviving probability of hosts $P_1(t)$ (dashed lines) and that of
parasites $P_2(t)$ 
(solid lines) with $\lambda_{01}=10$.
The lines from the top to the bottom correspond to
$\lambda_{12} = 5.9,6.3,6.7,7.1$, and 7.5.
We set $L=300$ in both (a) and (b).
}
%$\lambda_{12} = 5.5,5.9,6.3,6.7,7.1,7.5$.
\label{p10finite}
\end{figure}

Parasite-driven extinction may occur for large $\lambda_{12}$
\cite{haraguchi2000evolution,sato1994pathogen}.
Figure~\ref{p10finite}(b) shows the surviving probability of hosts
$P_1(t)$ and that of parasites $P_2(t)$ for some large values of
$\lambda_{12}$ and fixed values of $\lambda_{01}=10$ and $L=300$.
If $P_1(t)$ approaches zero rapidly, the parasite-driven extinction is considered
to have occurred.  If the transition from $S_{012}$ to the
parasite-driven extinction belongs to the DP universality class, $P_1(t)$ or
$P_2(t)$ should decay geometrically on the phase boundary and
exponentially for $\lambda_{12}$ slightly
larger than the critical value.  

%%%%%%%%%%%%%%%%%%%%%%%%%%%%%%%%%%%%%%%%%%%%%%%%%%%%%%%%%%%%%%%%%%%%%%%%%%%%%%%%%%%%%%%%%%%%%%%%%%%%

\begin{figure}[t]
\begin{center}
\includegraphics[width=7.6cm]{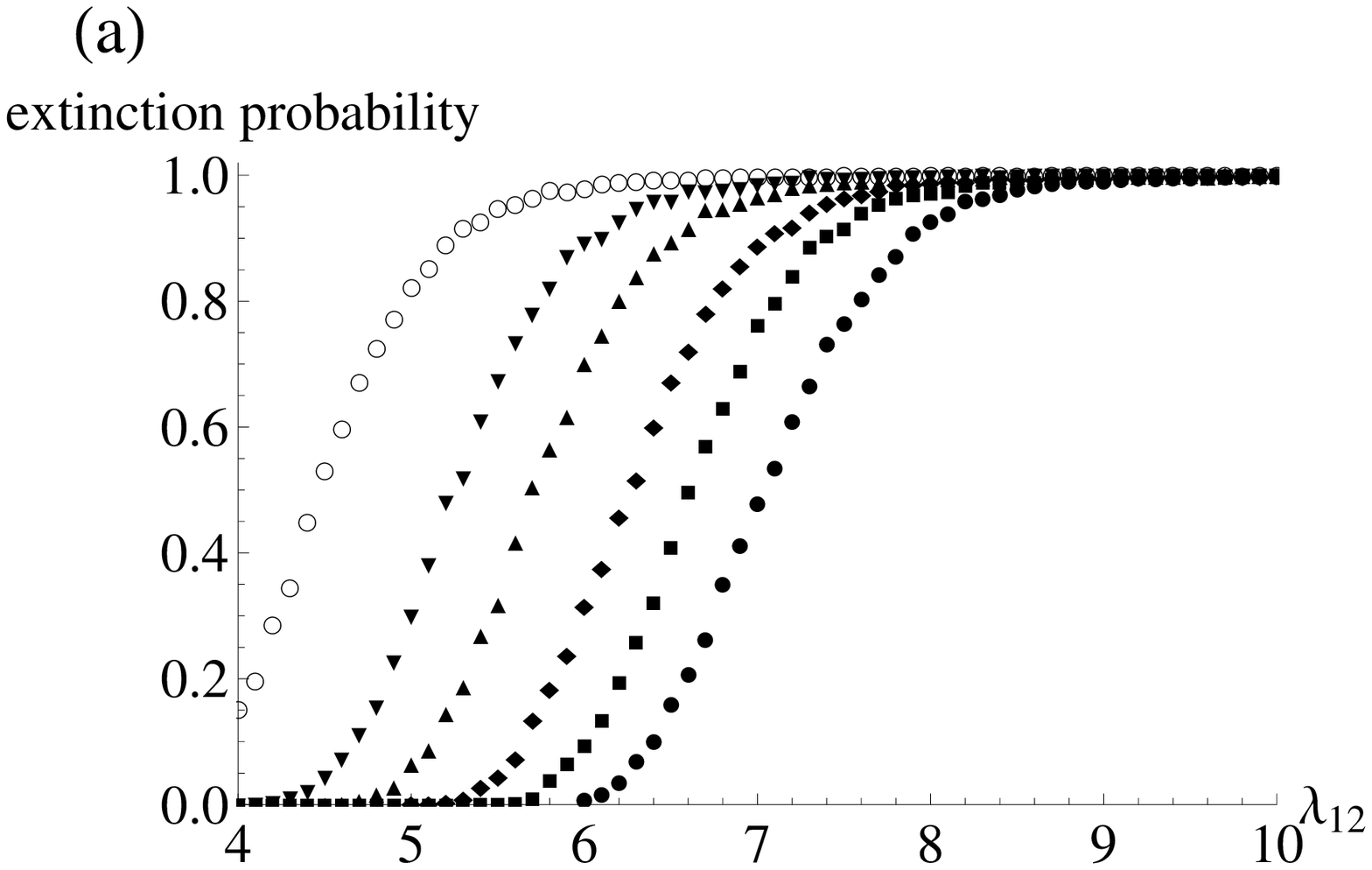}
\includegraphics[width=7cm]{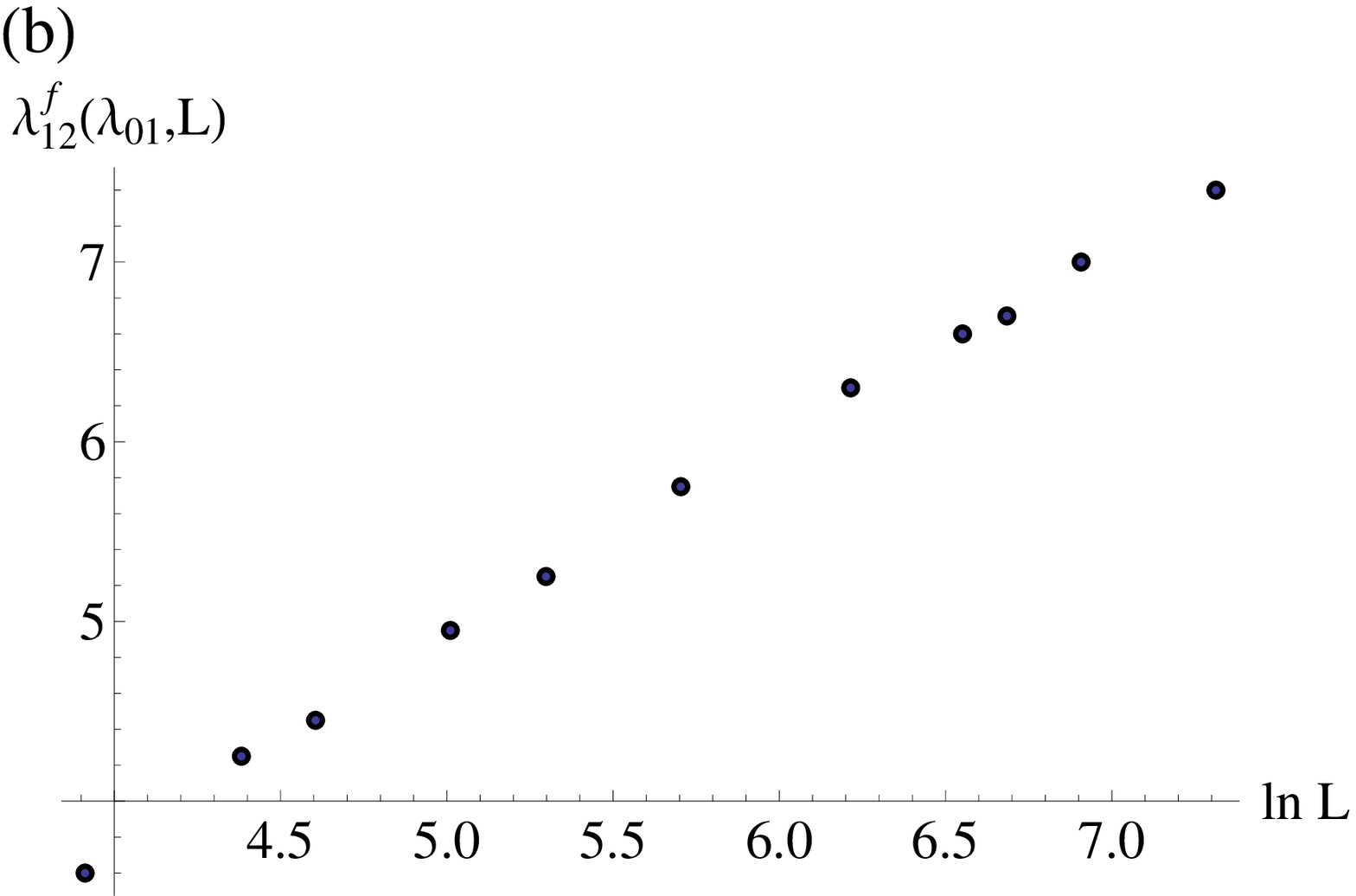}
\end{center}
\caption{
(a) Relationship between the extinction probability and 
$\lambda_{12}$ when
$L=100,200,300,500,700$, and 1000 (from left to right).
(b) Dependence of $\lambda_{12}^f(\lambda_{01}, L)$
on $L$.
%, at which the fraction of the PDEs takes half on the system 
%with $L=50,80,100,150,200,300,500,700,800,1000,1500$ (from left to right).
We set $\lambda_{01}=10$ in both (a) and (b).
The number of realizations for a given combination of
$\lambda_{12}$ and $L$ is equal to 2000.
}
\label{histogram}
\end{figure}

%%%%%%%%%%%%%%%%%%%%%%%%%%%%%%%%%%%%%%%%%%%%%%%%%%%%%%%%%%%%%%%%%%%%%%%%%%%%%%%%%%%%%%%%%%%%%%%%%%%%%%%%

However, Fig.~\ref{p10finite}(b)
indicates that this is not the case.  Whether extinction of hosts and
parasites occurs or not is determined at an early stage, 
where hosts are rapidly replaced by parasites, 
resulting in a rapid decrease in the number of hosts. 
If the hosts die out, the parasite-driven extinction takes place. 
In contrast, if hosts survive the initial stage, 
which occurs with
a low probability, the hosts recover from near extinction. 
In this case, hosts and parasites are likely to coexist
for long time. 
The value of $\lambda_{12}$ affects 
the probability that the hosts survive 
rather than the rates at which the number of hosts and parasites decay.

We state that the parasite-driven extinction is a finite size effect.
In order to confirm this statement, we measure the probability of
the parasite-driven extinction
as a function of linear lattice size $L$. Because the transient is short, as shown in Fig.~\ref{p10finite}(b),
we measure the fraction of realizations among 2000 realizations
in which both hosts and parasites are extinct
at $t=100$.
Figure~\ref{histogram}(a) shows the 
extinction probability for a range of values of $\lambda_{12}$ 
at $\lambda_{01}=10$ and $L=100,200,300,500,700$, and 1000.
The extinction probability indefinitely decreases with $L$. 
The value of $\lambda_{12}$
that has an extinction probability of $1/2$, denoted 
by $\lambda_{12}=\lambda_{12}^f(\lambda_{01}, L)$, 
is plotted against $L$ in Fig.~\ref{histogram}(b).
It is observed that $\lambda_{12}^f(\lambda_{01},L)\propto \ln L$.
Logarithmic scaling is also observed at other values of
$\lambda_{01}$. 
In Fig.~\ref{phasediagram}, we show $\lambda_{12}^f(\lambda_{01}, L)$ 
for some values of $\lambda_{01}$ and $L$ (dotted lines).

The results obtained in this section
indicate that the parameter region of parasite-driven
extinction indefinitely shrinks as $L$ increases.
This system-size dependence is distinct from
the dependence of the critical value on $L$ in the
usual phase transitions, which is convergent in the limit of infinite system size.

\section{DP transition on the $S_0$--$S_{01}$ boundary in the limit
$\lambda_{12} \to \infty$\label{sec:5}}

When $\lambda_{12}$ is sufficiently large, 
the mean field approximation predicts that
the system transits from $S_{0}$ to $S_{012}$
as $\lambda_{01}$ increases \cite{sato1994pathogen}.
The PA predicts that
the system transits from $S_{0}$ to $S_{01}$ and then to $S_{012}$
as $\lambda_{01}$ increases
\cite{ohtsuka2006phase}.
The i-PA predicts that the system is in $S_0$ irrespective of the value of $\lambda_{01}$ (see Figure 1 in \cite{haraguchi2000evolution}).
To analyze this apparent contradiction,
we carry out simulations in the limit 
$\lambda_{12}\to\infty$.

Irrespective of the value of $\lambda_{12}$,
it seems that
$\lambda_{01}$ must be larger than $\lambda_{01}^{\rm c}$ for hosts to 
survive. Therefore, we start by examining the case
$\lambda_{01} \approx \lambda_{01}^{\rm c}$.
When $\lambda_{01} \approx \lambda_{01}^{\rm c}$ and
$\lambda_{12}\to\infty$, a host adjacent to a parasite is instantaneously
invaded by the parasite. 
In such a case, if we start
numerical simulations
on the equal fraction of empty sites, hosts, and parasites, 
then the number of hosts, if they survive at all, becomes small
at the very beginning of a run.
For example,
the averaged number of hosts on the $300 \times 300$ square lattice
decreases from $30000$ to $\approx 60$ after a short time.
It may not be suitable to measure 
the decay of the expected number of hosts, 
which would be $\langle \rho_1 \rangle(t) \propto t^{-\delta}$ on the critical line; 
this is because such a measurement necessitates the existence of a sufficiently large number of hosts at the beginning of a run.

Another numerical method for estimating
the transition point and critical exponents 
is to measure the time courses of
the system starting from an almost absorbing configuration 
\cite{grassberger1979reggeon}.
For example, 
we observe the power law behavior
of the surviving probability, the number of active sites, and the mean spreading at the transition point, if we run the CP 
starting from a single active site.
Therefore, we assume that the initial configuration of the HP model
contains just one host. 
The other sites are either empty or parasites with a probability of 0.5.

%%%%%%%%%%%%%%%%%%%%%%%%%%%%%%%%%%%%%%%%%%%%%%%%%%%%%%%%%%%%%%%%%%%%%%%%%%%%%%%%%%%%%%%%%%%%%%%%%%%%

\begin{figure}[!t]
\begin{center}
\includegraphics[width=7cm]{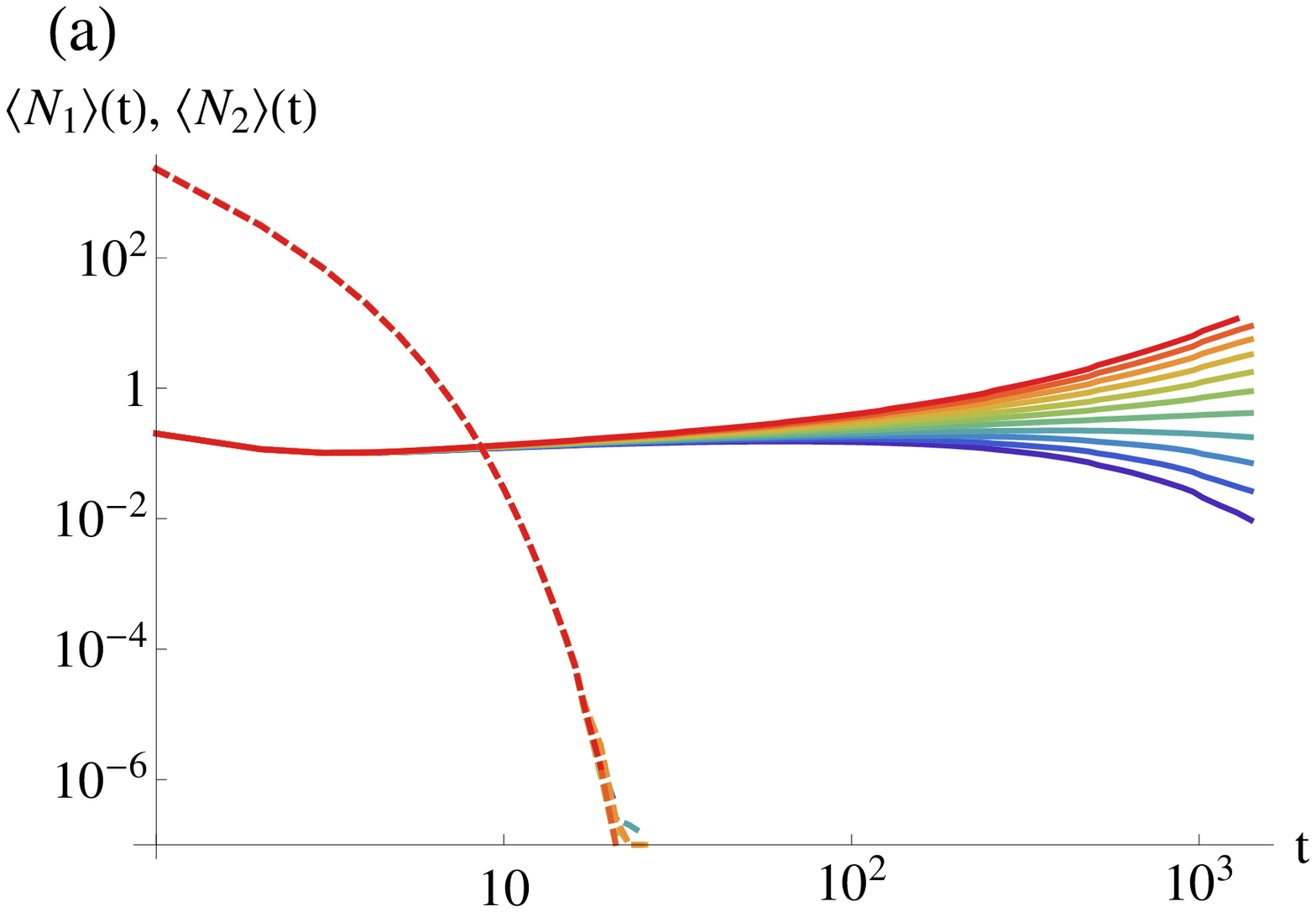}
\includegraphics[width=7cm]{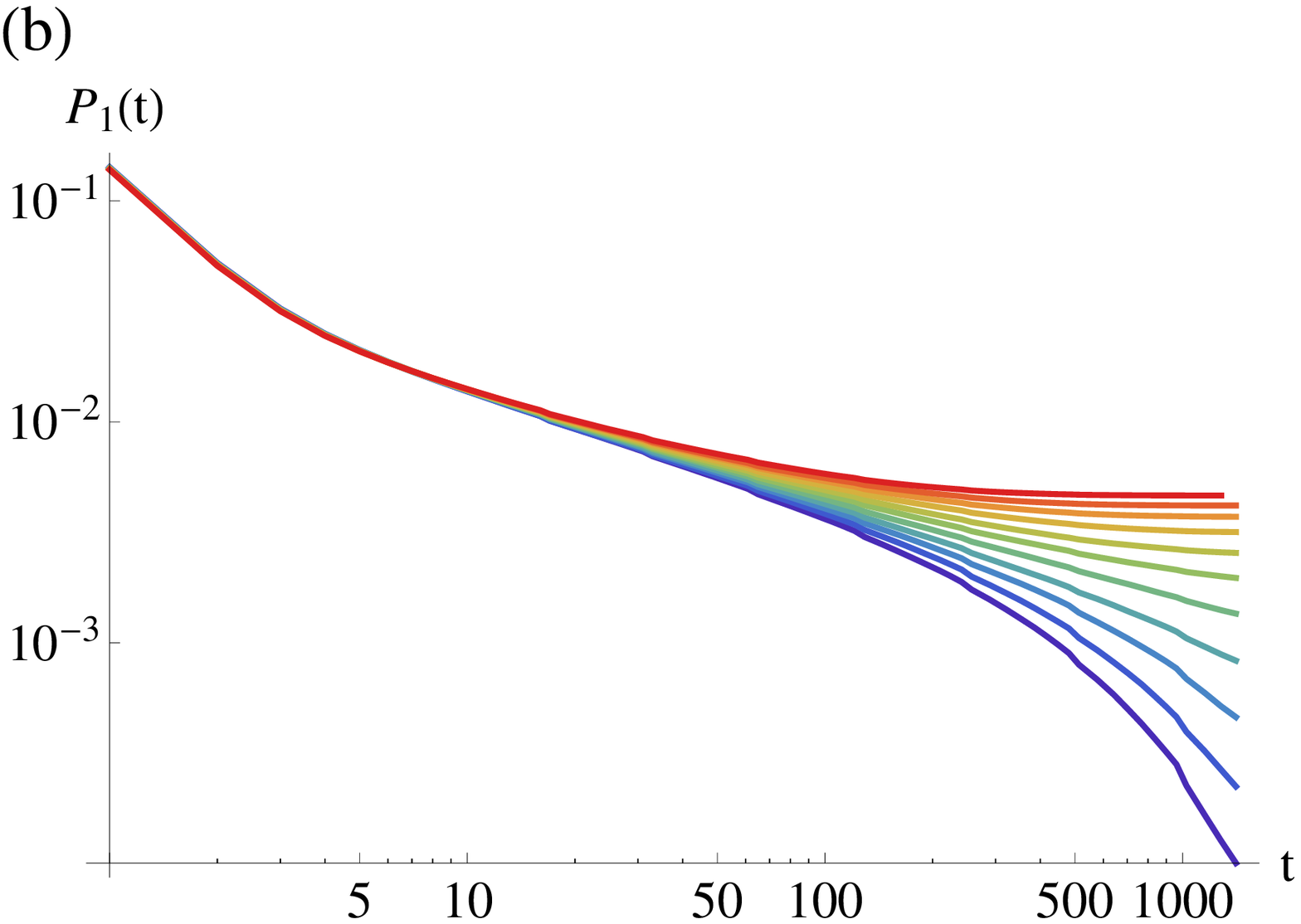}
\includegraphics[width=7cm]{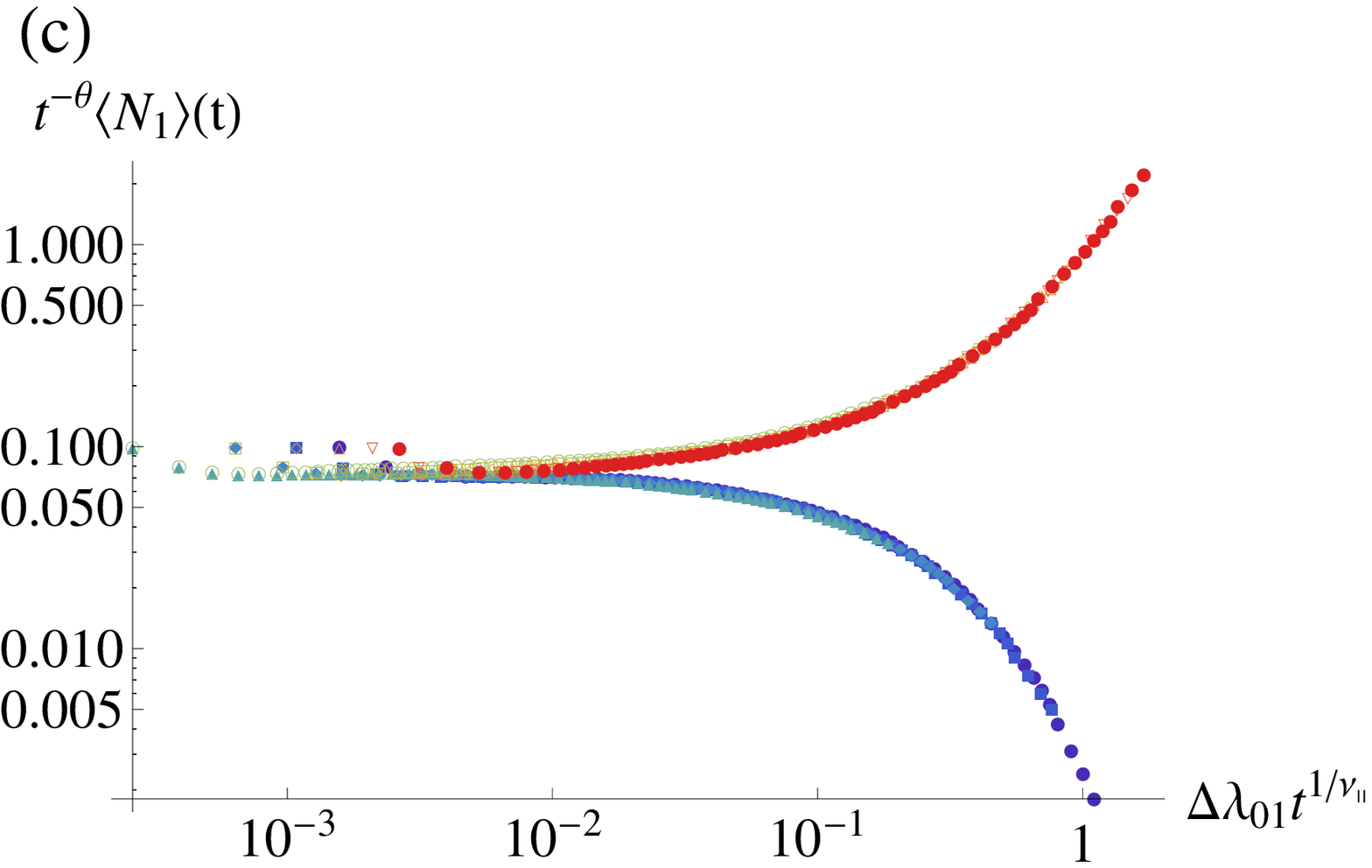}
\includegraphics[width=7cm]{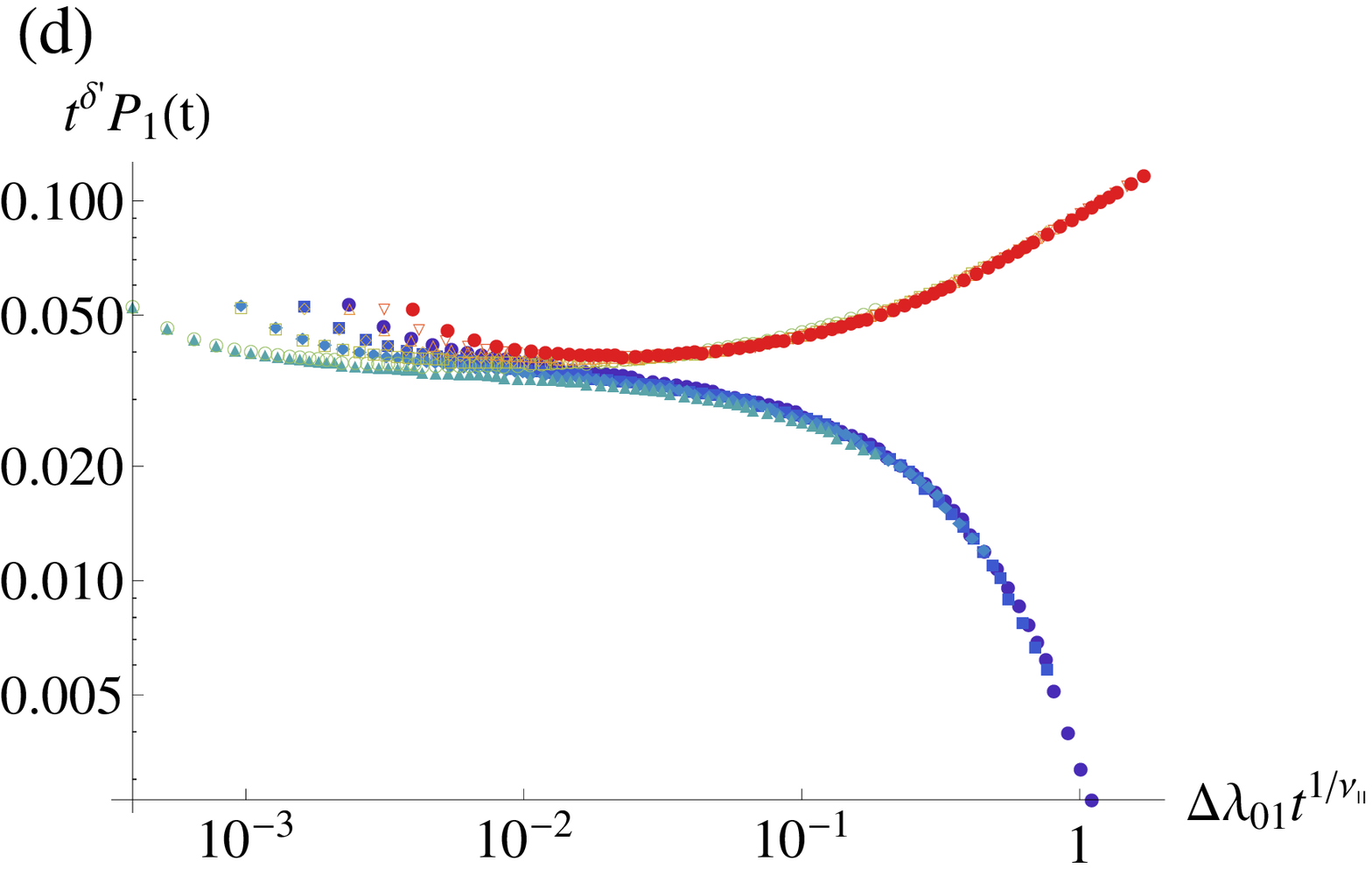}
\end{center}
\caption{Transitions at $\lambda_{01} \approx \lambda_{01}^{\rm c}$ and
$\lambda_{12}\to\infty$.
(a) Time courses of $\langle N_1 \rangle (t)$ (solid lines) and $\langle N_2 \rangle (t)$ (dashed lines that almost overlap each other).
(b) Surviving probability of hosts $P_1(t)$.
(c) Dynamic scaling (Eq.~\eqref{N}) for the data shown in (a).
(d) Dynamic scaling (Eq.~\eqref{P}) for the data shown in (b). 
The lines correspond to $\lambda_{01}=0.4082, 0.4092, \ldots$, and 0.4182
from the
bottom to the top.
The number of realizations for a given $\lambda_{01}$ is equal to $10^7$.
}
\label{p21infty}
\end{figure}

%%%%%%%%%%%%%%%%%%%%%%%%%%%%%%%%%%%%%%%%%%%%%%%%%%%%%%%%%%%%%%%%%%%%%%%%%%%%%%%%%%%%%%%%%%%%%%%%%%%%

With this one-host configuration, 
the mean number of hosts follows the power law
$\langle N_1 \rangle (t) \propto t^{\theta}$
at $\lambda_{01} \approx \lambda_{01}^{\rm c}$,
as shown by the solid lines in Fig.~\ref{p21infty}(a).
On the other hand, parasites rapidly become extinct (dashed line).
The surviving probability of hosts 
also follows the approximate power law $P_1(t) \propto t^{-\delta'}$
in the same parameter range (Fig.~\ref{p21infty}(b)).

At $\lambda_{01} \approx \lambda_{01}^{\rm c}$,
we adopt 
the dynamic scaling ansatz \cite{marro1999nonequilibrium} represented by
\begin{equation}
\langle N_1 \rangle (t) \approx t^{\theta} \tilde{N_1}
\left(\Delta \lambda_{01} t^{1/\nu_{||}}, \frac{t^{d/z}}{N}\right),  
 \label{N} 
 \end{equation}
\begin{equation}
P_1(t) \approx t^{-\delta'} \tilde{P_1}\left(
\Delta \lambda_{01} t^{1/\nu_{||}}, \frac{t^{d/z}}{N}\right), \label{P}
 \end{equation}
where
\begin{equation}
\Delta \lambda_{01}=\lambda_{01}-\lambda_{01}^{\rm c}.
\end{equation}
This dynamic scaling ansatz explains the data shown in Figs.~\ref{p21infty}(a) and ~\ref{p21infty}(b), respectively.  The fitting results with the DP
exponents $\theta \approx 0.229$ 
and $\delta'=\delta \approx 0.451$ \cite{marro1999nonequilibrium} 
(Figs.~\ref{p21infty}(c) and ~\ref{p21infty}(d)) suggest that the
transition from $S_0$ to $S_{01}$ at $\lambda_{01} = \lambda_{01}^{\rm
  c}$ and $\lambda_{12}\to\infty$ is of the DP type. 
We consider that this phase transition is independent of the value of $\lambda_{12}$. 
This result qualitatively agrees with that obtained from the PA but not that obtained from the i-PA.  

%%%%%%%%%%%%%%%%%%%%%%%%%%%%%%%%%%%%%%%%%%%%%%%%%%%%%%%%%%%%%%%%%%%%%%%%%%

\begin{figure}[!t]
\begin{center}
\includegraphics[width=7cm]{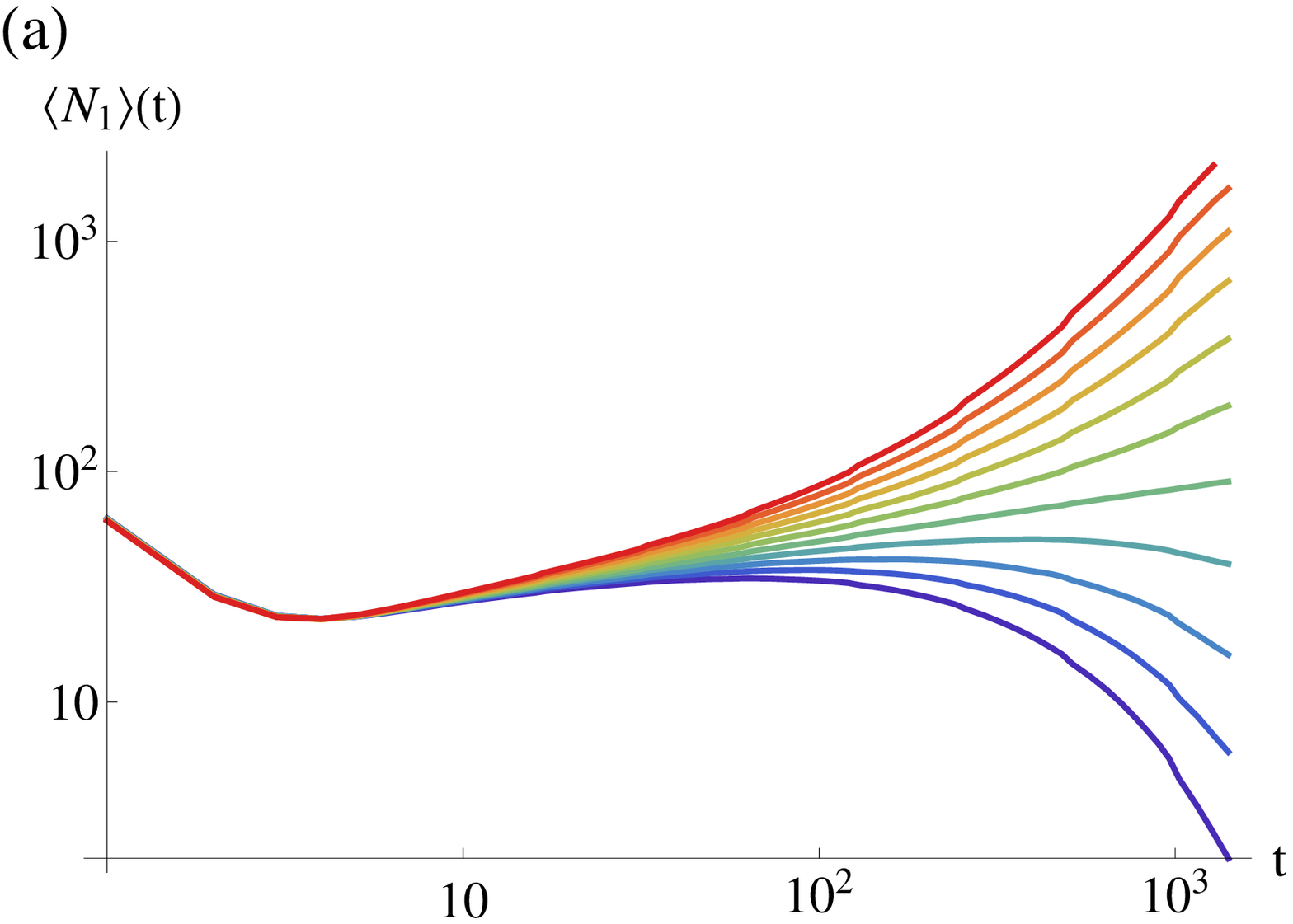}
\includegraphics[width=7cm]{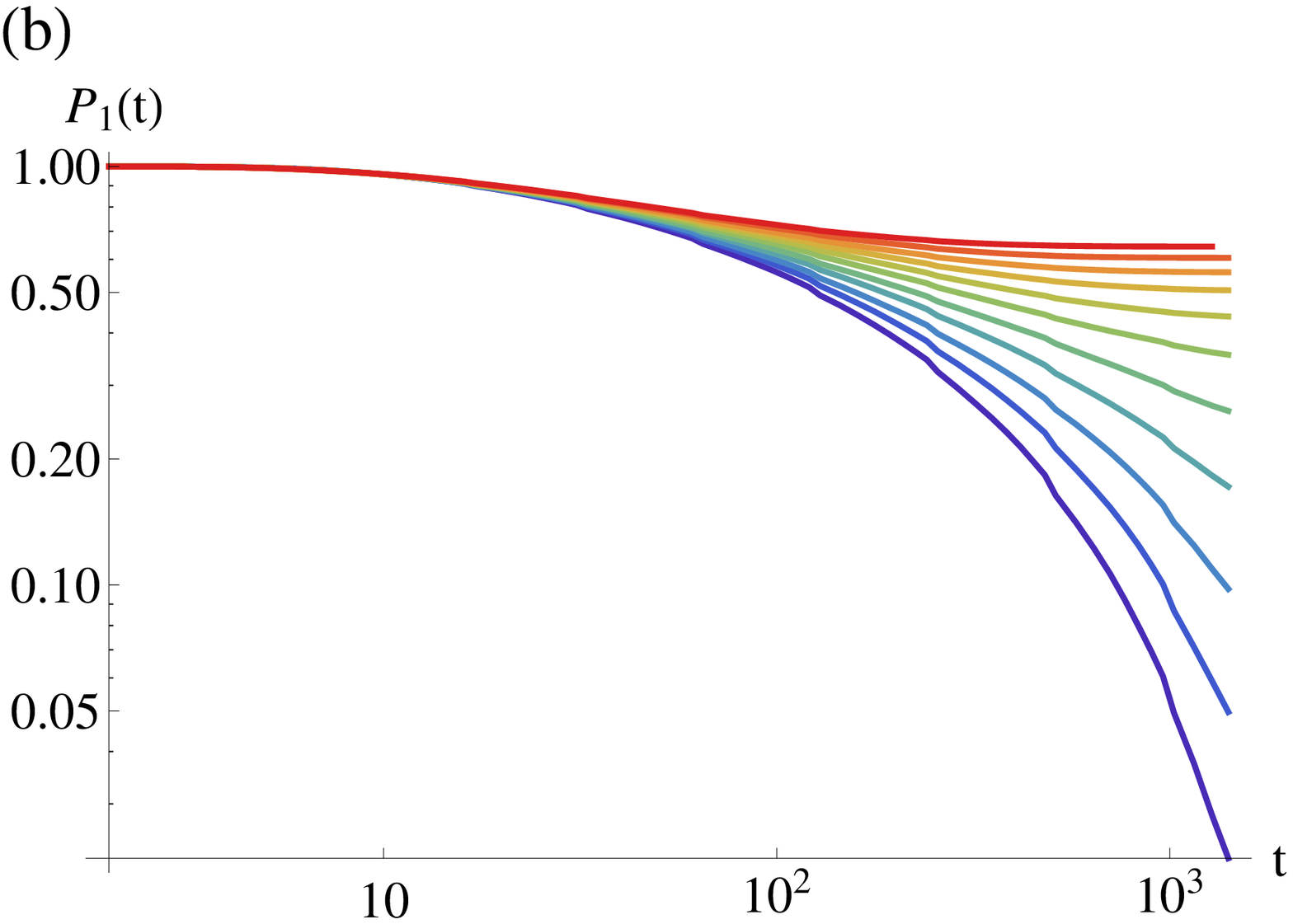}
\includegraphics[width=7cm]{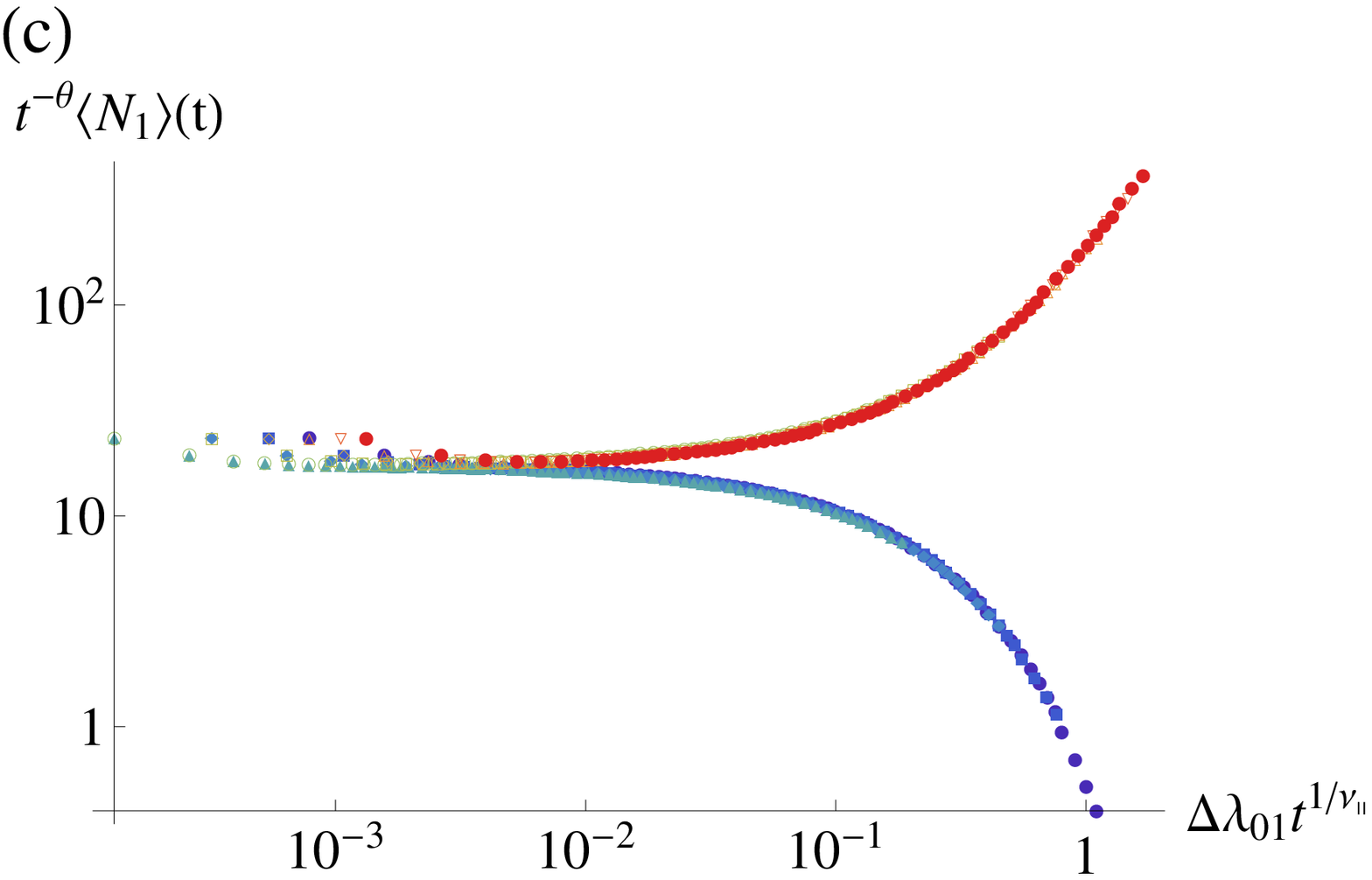}
\includegraphics[width=7cm]{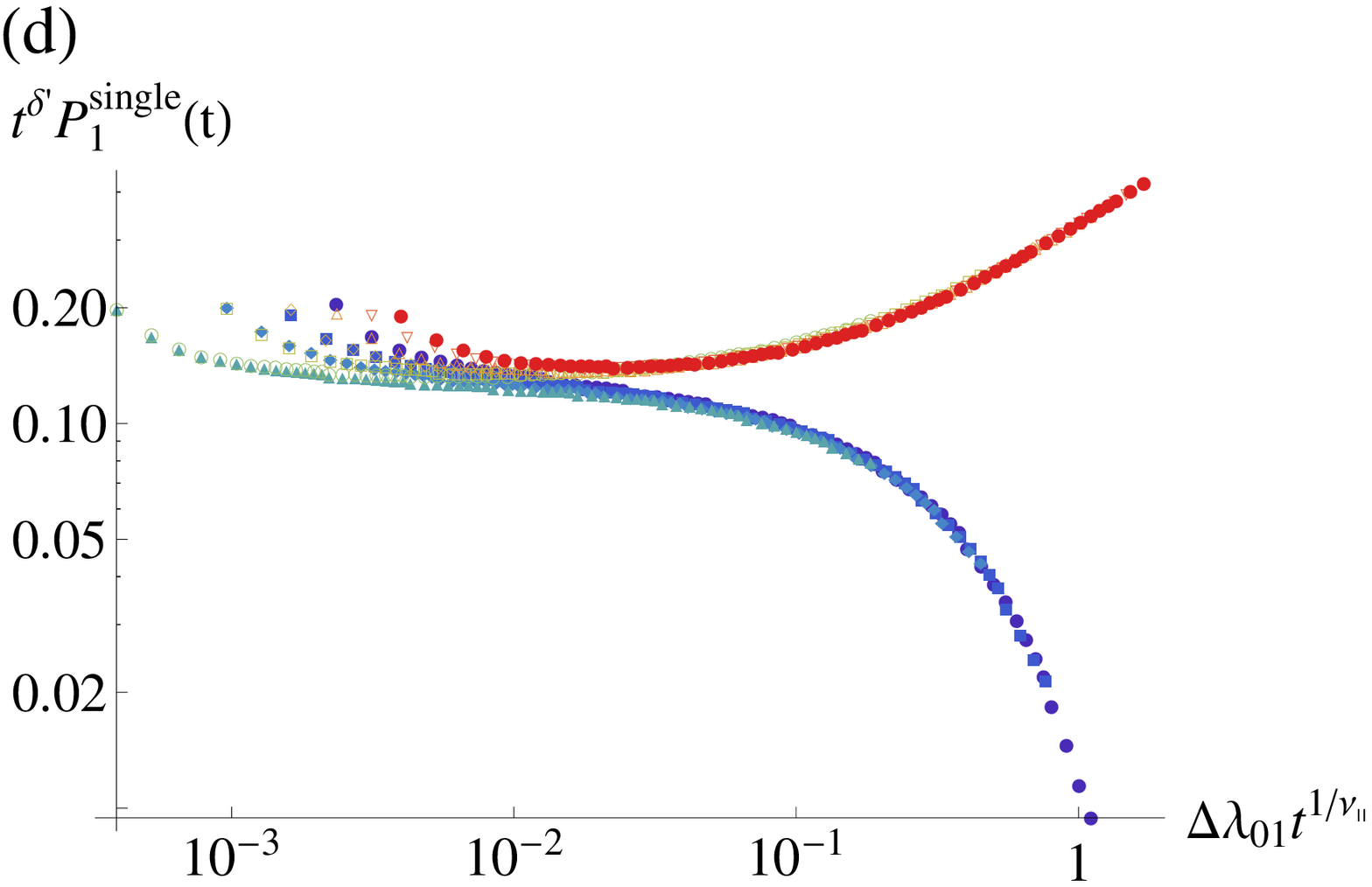}
\end{center}
\caption{
(a) Time courses of $\langle N_1 \rangle (t)$ when $\lambda_{01} \approx \lambda_{01}^{\rm c}$.
(b) Surviving probability of hosts $P_1(t)$.
(c) Dynamic scaling (Eq.~\eqref{N}) for the data shown in (a).
(d) Dynamic scaling with Eqs.~\eqref{P} and~\eqref{single} applied to
the data shown in (b). 
The lines correspond to $\lambda_{01}=0.40821, 0.40921, \ldots$, and 
0.41821 from the bottom to the top.
The number of realizations for a given $\lambda_{01}$ is equal to $10^7$.
}
\label{p21inftyrand}
\end{figure}

%%%%%%%%%%%%%%%%%%%%%%%%%%%%%%%%%%%%%%%%%%%%%%%%%%%%%%%%%%%%%%%%%%%%%%%%%%

With the random initial configuration,
we observe $\langle N_1 \rangle (t)$ and $P_1(t)$ instead of $\langle \rho_1 \rangle (t)$ 
and obtain the same results as those shown in Fig.~\ref{p21infty}.
$\langle N_1 \rangle (t)$ and
$P_1(t)$ decay geometrically
at $\lambda_{01} \approx \lambda_{01}^c$,
as shown in Fig.~\ref{p21inftyrand}(a) and \ref{p21inftyrand}(b),
respectively. 
The dynamic scaling (Eq.~\eqref{N}) with the DP exponents
fits $\langle N_1 \rangle (t)$ shown in 
Fig.~\ref{p21inftyrand}(a) well (Fig.~\ref{p21inftyrand}(c)).
On the other hand, 
dynamic scaling of $P_1(t)$ (Eq.~\eqref{P}) fails because the number of surviving hosts after a short time is greater than one.
To circumvent this case, we assume that
the surviving hosts are located away from each other
and grow independently on the lattice.
We denote the surviving probability of a specified host by
$P_1^{\rm single}(t)$.
Then, we approximate $P_1(t)$ as
\begin{equation}
P_1(t) \approx 1-(1-P_1^{\rm single}(t))^n,
\end{equation}
that is,
\begin{equation}
P_1^{\rm single}(t) \approx 1-(1-P_1(t))^{1/n}, \label{single}
\end{equation}
where $n$ is the mean number of surviving hosts after a short time.
By replacing $P_1(t)$ in Eq.~\eqref{P} by $P_1^{\rm single}(t)$ 
and using the DP critical exponents,
we obtain a reasonable scaling,
as shown in Fig.~\ref{p21inftyrand}(d).

%%%%%%%%%%%%%%%%%%%%%%%%%%%%%%%%%%%%%%%%%%%%%%%%%%%%%%%%%%%%%%%%%%%%%%%%%%%%%%%%%%%%%%%%%%%%%%%%%%%%

\begin{figure}[!t]
\begin{center}
\includegraphics[width=7cm]{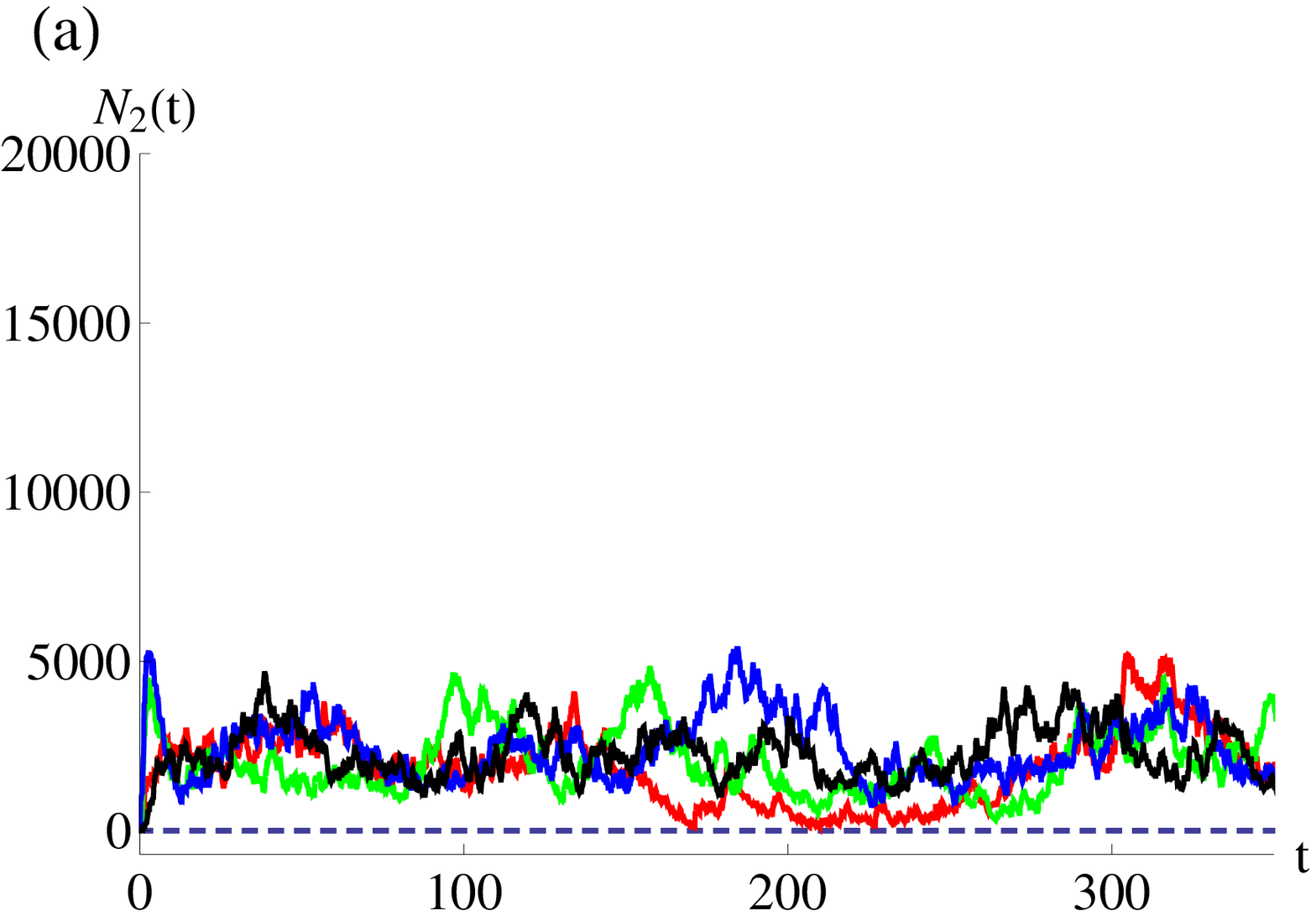}
\includegraphics[width=7cm]{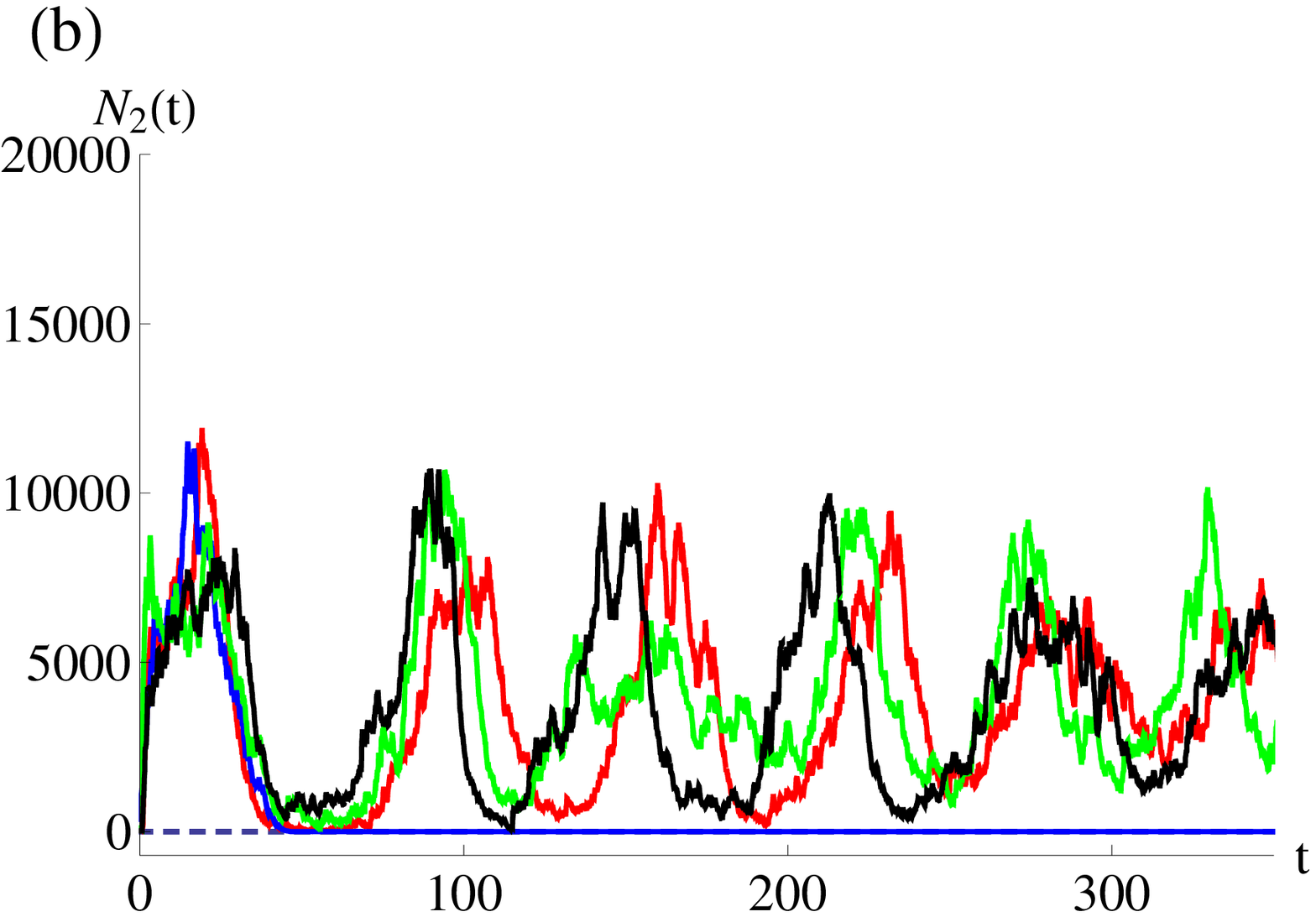}
\includegraphics[width=7cm]{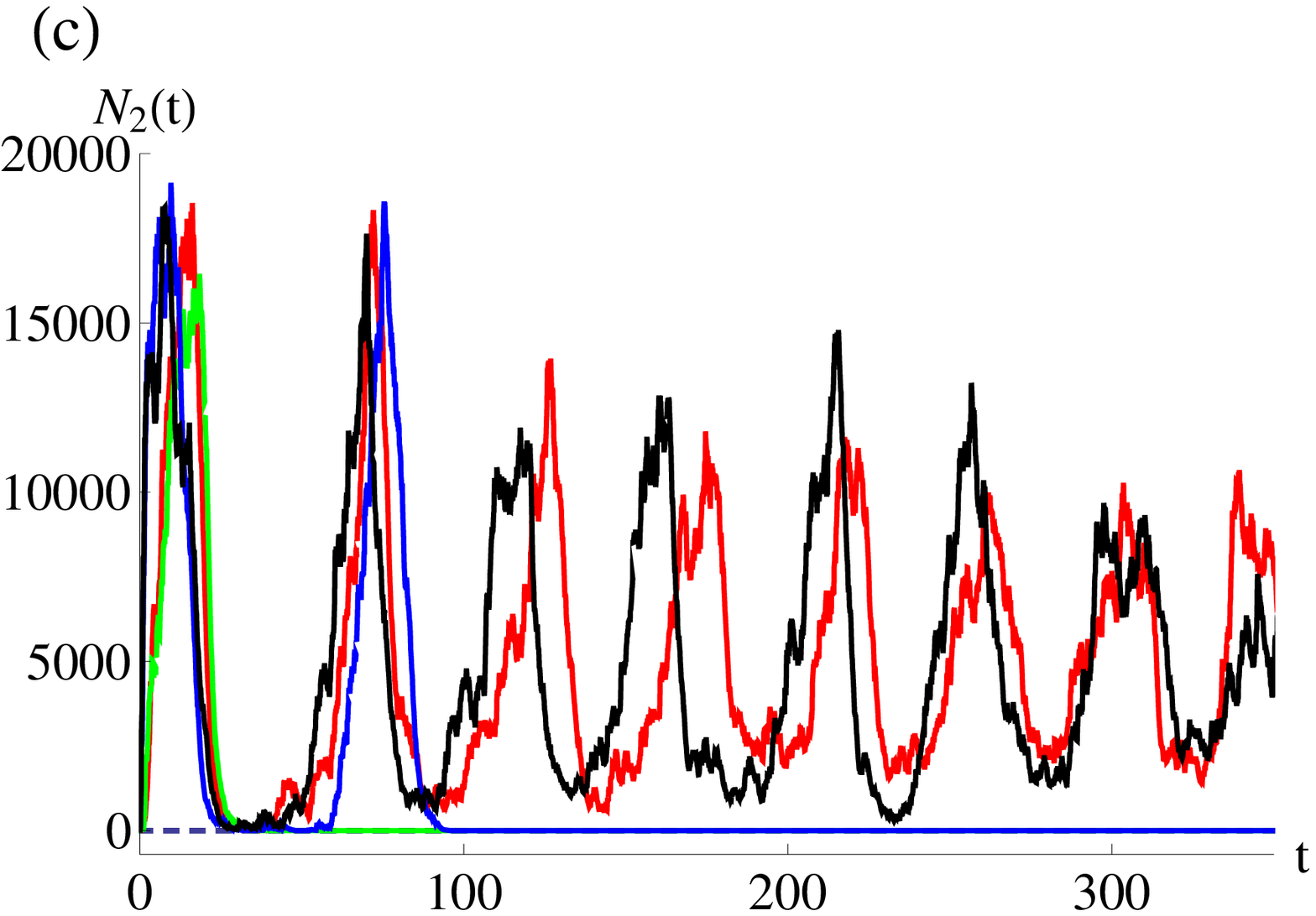}
\end{center}
\caption{Time courses of $N_2 (t)$ when
$\lambda_{01}=$  (a) 0.515, (b) $0.530$, and (c) $0.545$.
We set $\lambda_{12}\to\infty$ and $L=700$.
Each colored line represents a single run, and
the results for 4 runs are shown in each panel.
}
\label{p21inftyS012}
\end{figure}

\begin{figure}
\begin{center}
\includegraphics[width=7cm]{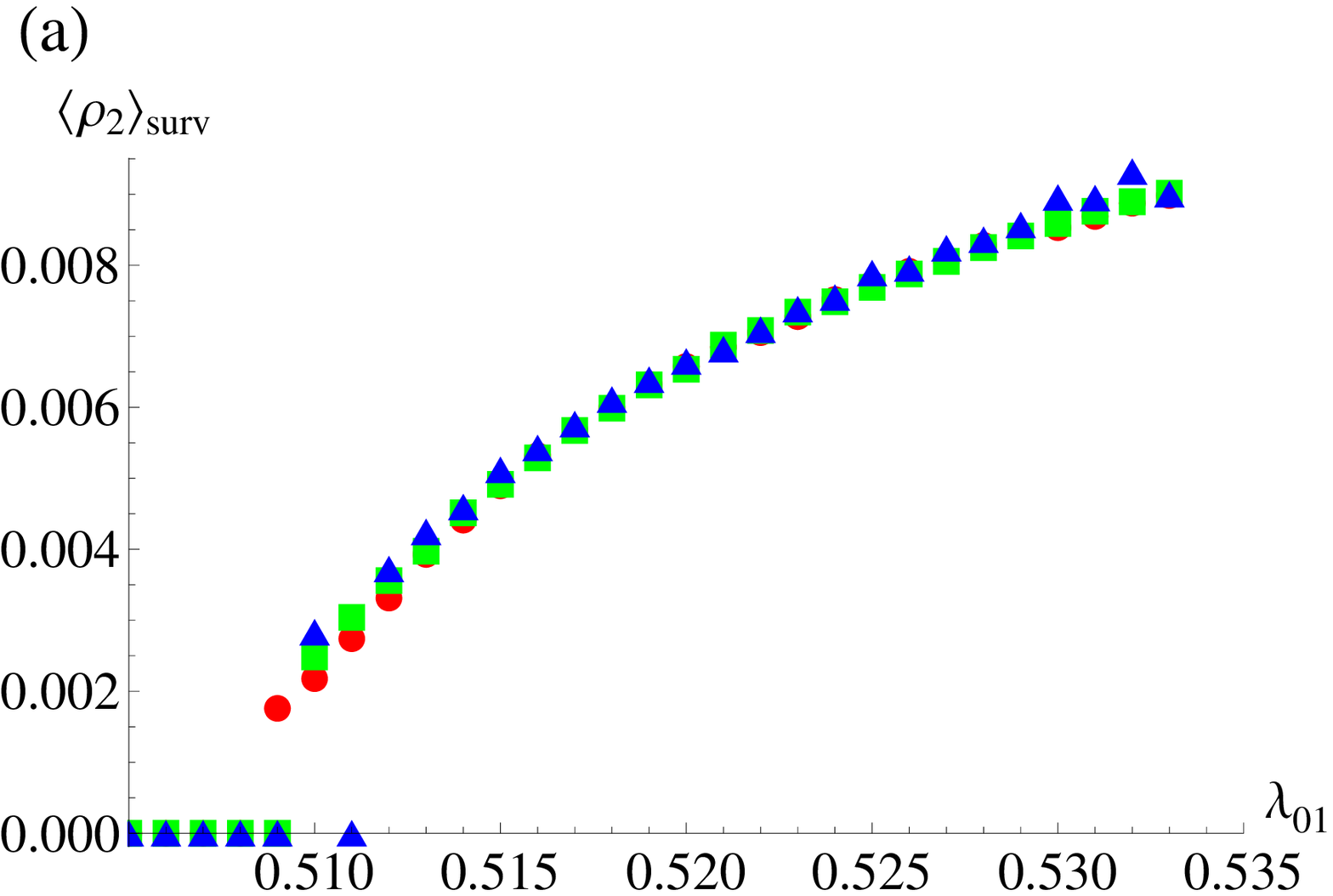}
\includegraphics[width=7.2cm]{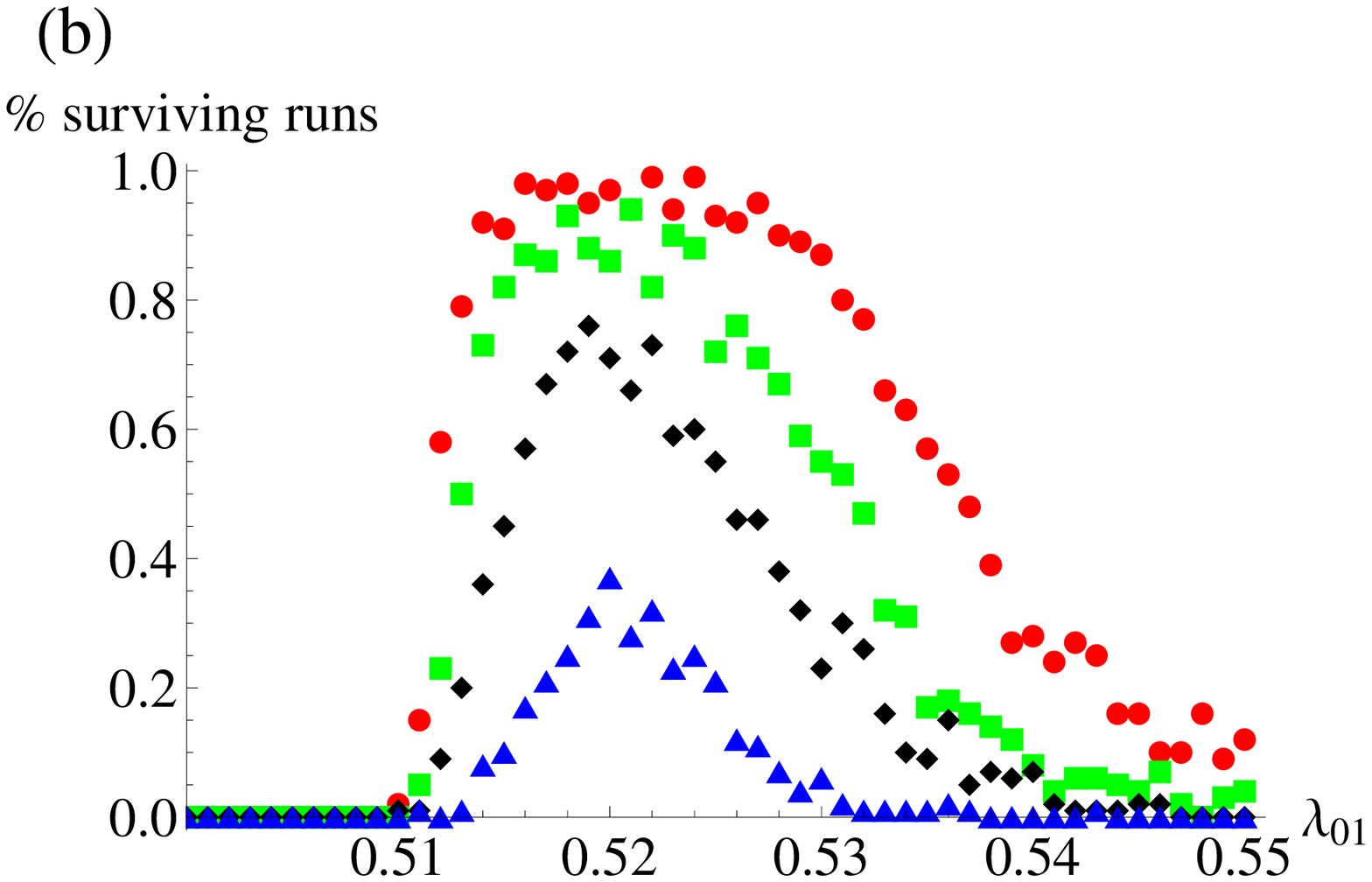}
\end{center}
\caption{
(a) Stationary parasite density $\langle \rho_2 \rangle_{\rm surv}$ averaged
over the surviving runs in the limit $\lambda_{12} \to \infty$. 
(b) Fraction of the surviving runs.
We set $L=500$ (triangles), $600$ (diamonds; only in (b)), 
$700$ (squares), and $900$ (circles).
The number of realizations for a given combination of
$\lambda_{01}$ and $L$ is equal to 100.
}
\label{p21inftyS012Histo}
\end{figure}

\begin{figure}
\begin{center}
\includegraphics[width=7cm]{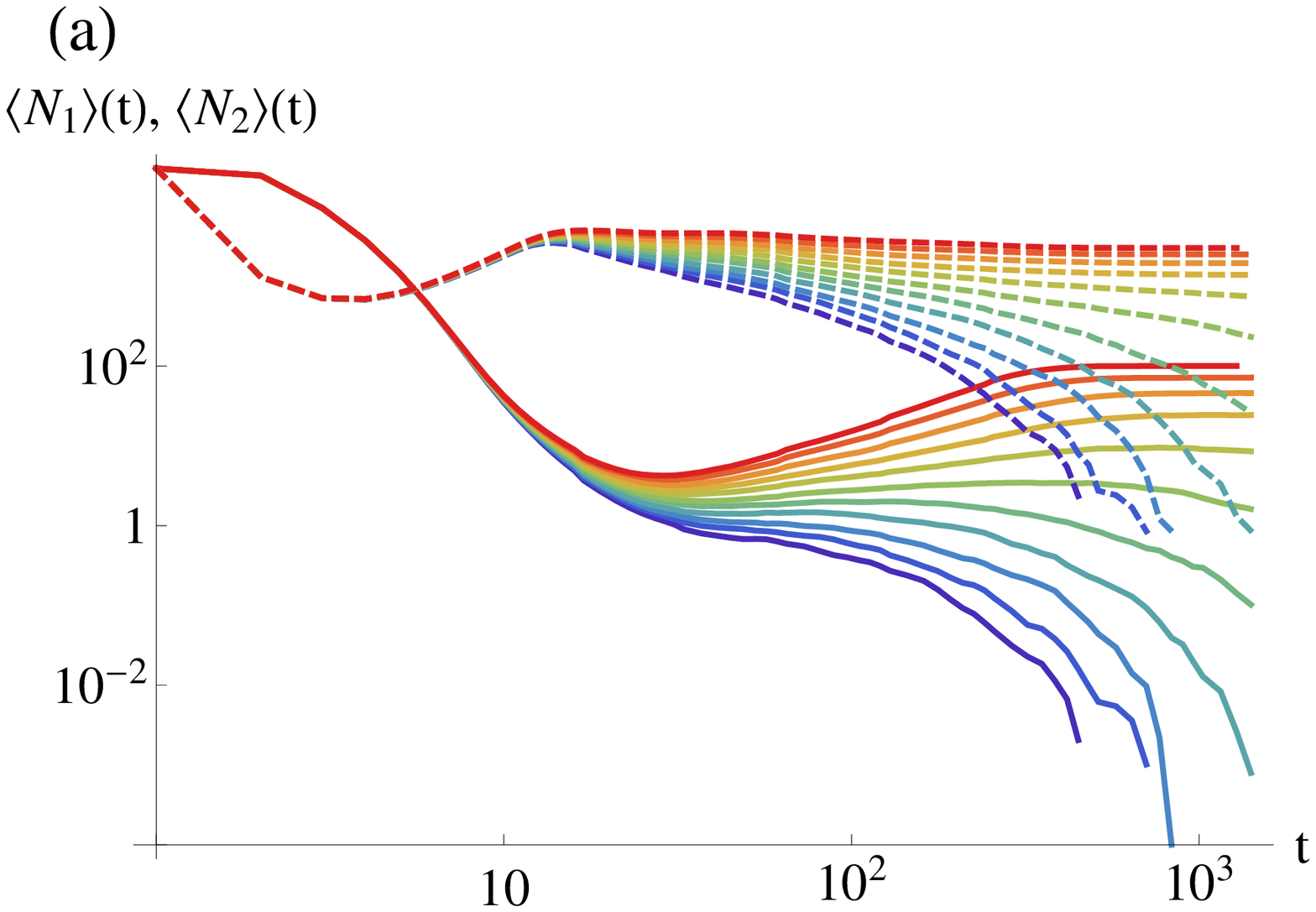}
\includegraphics[width=7cm]{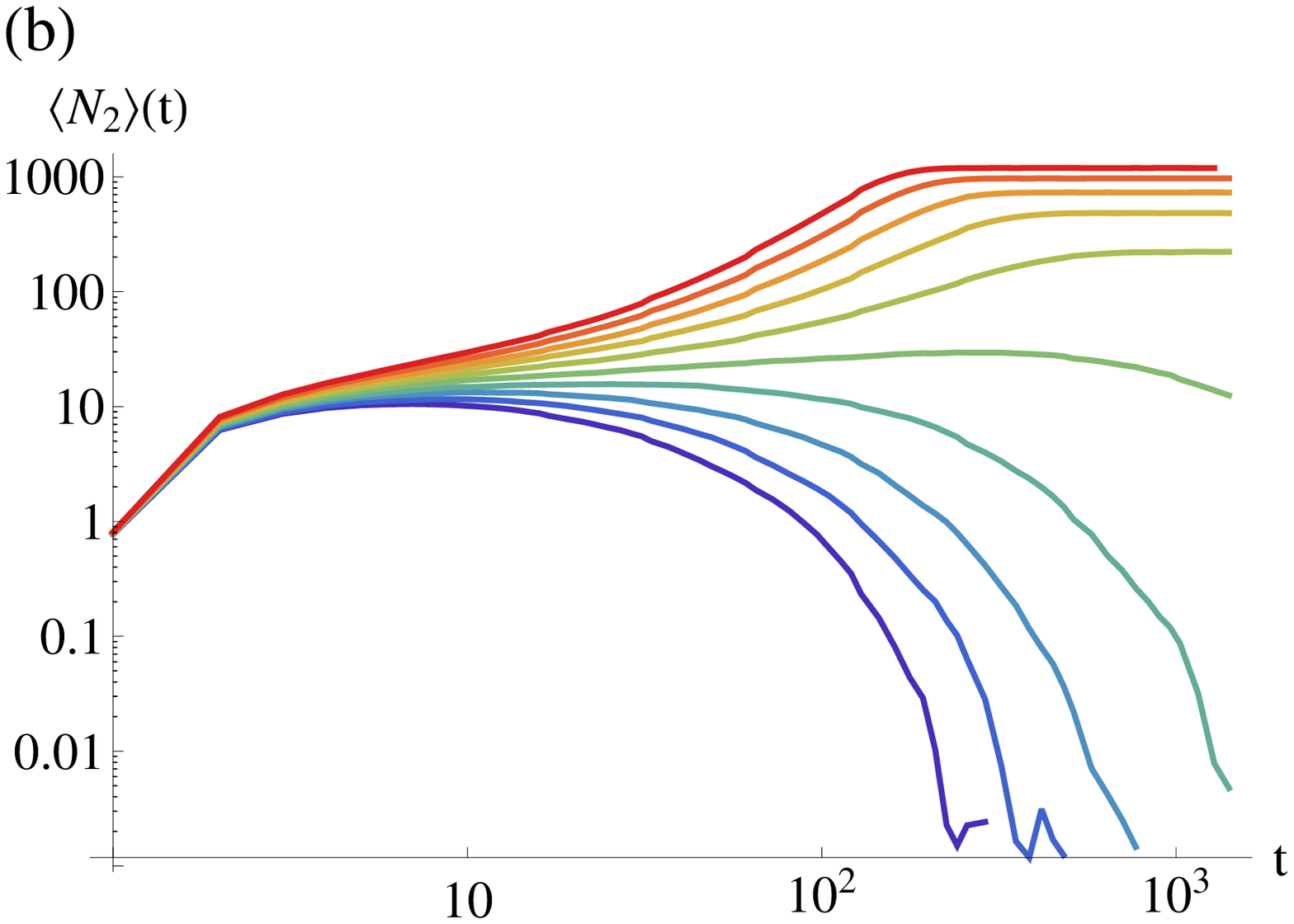}
\includegraphics[width=7cm]{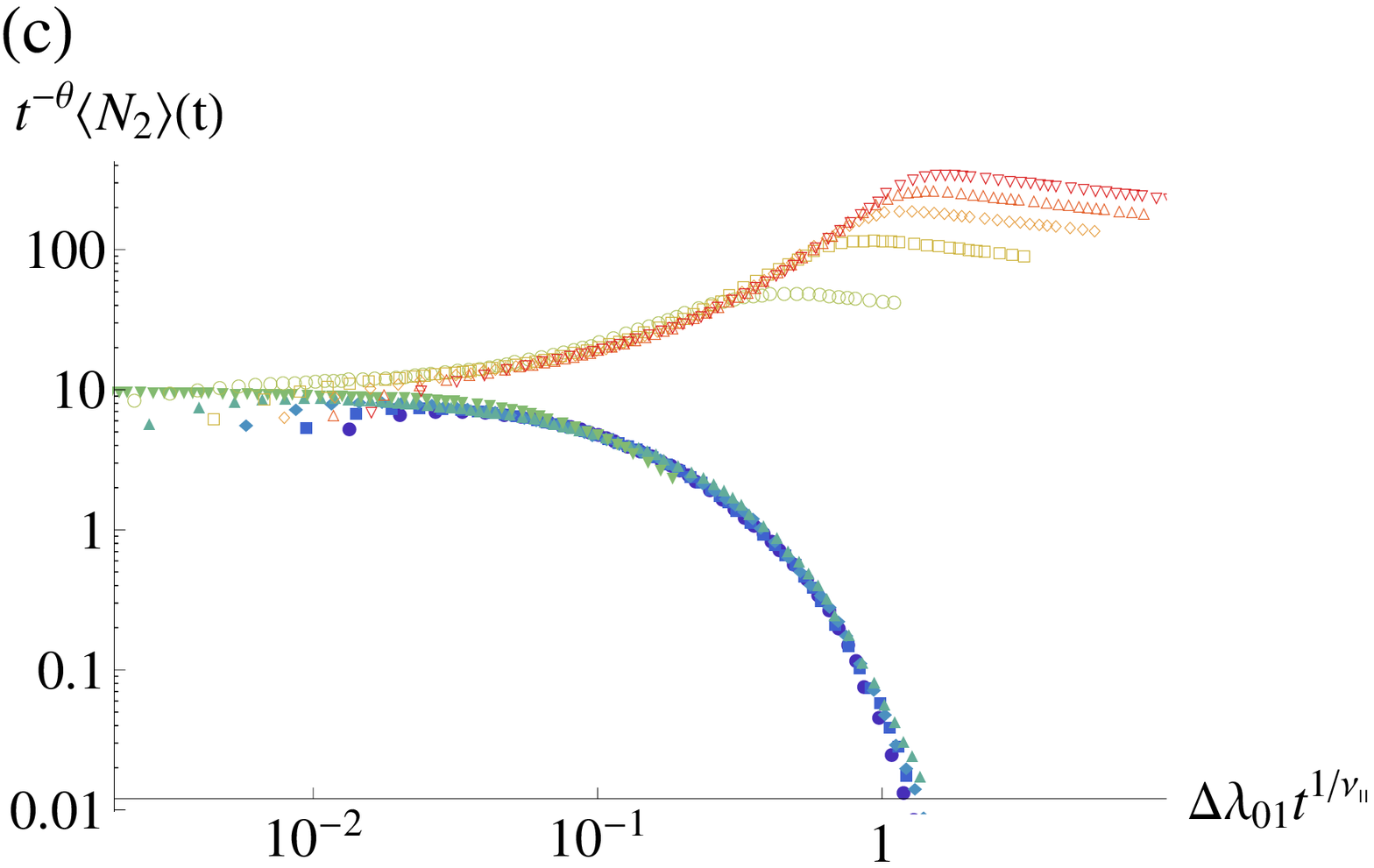}
\end{center}
\caption{
(a) Time courses of $\langle N_1 \rangle (t)$ (dashed lines) and $\langle N_2 \rangle (t)$ (solid lines) for the random initial configuration.
The lines from the bottom to the top correspond to
$\lambda_{01}=0.58, 0.582, \ldots$, and 0.60. 
The number of realizations for a given $\lambda_{01}$ is equal to 50000.
(b) Time courses of $\langle N_2 \rangle (t)$
for the modified initial configuration.
The lines correspond to $\lambda_{01}=0.57, 0.575, \ldots$, and 0.615
from the bottom to the top. 
The number of realizations for a given $\lambda_{01}$ is equal to 20000.
We set $\lambda_{12}=4$ and $L=300$ in both (a) and (b).
(c) Dynamic scaling for the data shown in (b). 
As the scaling function, we use Eq.~\eqref{N} with $\langle N_1 \rangle (t)$ replaced
by $\langle N_2 \rangle (t)$. 
}
\label{additionalFig}
\end{figure}

%%%%%%%%%%%%%%%%%%%%%%%%%%%%%%%%%%%%%%%%%%%%%%%%%%%%%%%%%%%%%%%%%%%%%%%%%%%%%%%%%%%%%%%%%%%%%%%%%%%%%%%%

\section{$S_{012}$ phase in the limit $\lambda_{12} \to \infty$\label{sec:6}}

When $\lambda_{12}\to\infty$, 
either the random initial configuration or
the one-host configuration yields $S_0$ or $S_{01}$, but 
not $S_{012}$, for any value of
$\lambda_{01}$. 
This remains the case for at least up to 
$L=1000$.
The apparent absence of $S_{012}$ may be because
there are initially too many parasites.
In the case of a large $\lambda_{12}$, parasites replace
hosts in a short time, which is likely to lead to the extinction of
the parasite.

To examine the possibility of $S_{012}$ at $\lambda_{12}\to\infty$,
we adopt the one-parasite configuration, where
the remaining sites are either empty or occupied by the host
with a probability of 0.5. 
With this initial configuration, we find that
both hosts and parasites can survive when $L$ is large and
$\lambda_{01}$ is within a certain range.
When $L \lesssim 400$, neither hosts nor parasites survive.

Time courses of the number of parasites are shown 
in Fig.~\ref{p21inftyS012} for $L=700$ and three values of $\lambda_{01}$.
As $\lambda_{01}$ increases within this range,
the basal number of parasites in a short run increases, but the amplitude
of the damped oscillation in the number of parasites also
increases.
If $\lambda_{01}$ is sufficiently large, the amplitude of the oscillation is so
large that the parasites are likely to disappear
in the first cycle of the oscillation
(Fig.~\ref{p21inftyS012}(c)), whereas the basal number of parasites is larger than that in the case of a smaller $\lambda_{01}$
(e.g., Fig.~\ref{p21inftyS012}(a)).
We remark that,
for related spatial stochastic processes,
sustainable oscillations \cite{itoh1994stochastic,morita2006undamped}
and absorption to the unanimity state owing to the blowing out of oscillations
\cite{szabo2004rock} were reported as finite size effects.

The stationary density of the parasites 
averaged over the surviving runs, denoted by $\langle
\rho_2\rangle_{\rm surv}$, is shown for some large values of $L$
in Fig.~\ref{p21inftyS012Histo}(a).
Here $\langle . \rangle_{\rm surv}$ indicates the average over realizations in
which parasites survive after a transient of length $1500$. 
We observe that $\langle \rho_2\rangle_{\rm surv}$ is
positive for $\lambda_{01} \gtrsim 0.509$
and converges to a certain value for $\lambda_{01} \gtrsim 0.518$.
We did not determine the transition point and the critical exponents 
by a scaling argument for $\langle \rho_2 \rangle_{\rm surv}$ in terms of
$\lambda_{01}$ because $\langle \rho_2 \rangle_{\rm surv}$ is too small 
for $\lambda_{01} \approx 0.509$. 
To support the existence of the $S_{012}$ phase in the limit $L\to\infty$, 
we measure the fraction of surviving runs for various system sizes.
As shown in Fig.~\ref{p21inftyS012Histo}(b), 
the fraction of surviving runs increases with $L$ for $\lambda_{01} \gtrsim 0.509$.
This result supports the fact that $S_{012}$ exists for $\lambda_{01} \gtrsim 0.509$ in the limit $L \to \infty$.
As $\lambda_{01}$ increases even further (i.e., $\lambda_{01} \gtrsim 0.524$), 
the fraction of surviving runs decreases.
The parasite-driven extinction for a finite system size gets eminent in this range of $\lambda_{01}$;
this parasite-driven extinction is caused by the increasing magnitude of damped oscillations. 
Similar to the results shown in Sec.~\ref{sec:4},
the parameter region for the
parasite-driven extinction depends on the system size and
is likely to disappear in the limit $L\to\infty$.
We also observed that the results in the case of finite $\lambda_{12} \gtrsim 2$
are qualitatively the same as those in the case of $\lambda_{12}=\infty$.

Finally, we examine the $S_{01}$--$S_{012}$ transition line 
for large $\lambda_{12}$. 
In this case, we do not obtain a data collapse by 
the dynamic scaling based on the relaxation of the system,
as shown in Fig.~\ref{additionalFig}(a) for $\lambda_{12}=4$.
Therefore, we attempt the dynamic scaling for the parasites 
in the manner similar to that employed in Sec.~\ref{sec:5}. 
Consider the neighborhood of the $S_{01}$--$S_{012}$ transition point for a large fixed
$\lambda_{12}$.
With the one-parasite configuration,
a parasite would quickly invade hosts at an early stage.
In this case, the growth rate of the parasite is fairly insensitive to $\lambda_{01}$.
Therefore, the scaling argument would not apply.

To avoid such an initial growth of parasites 
and obtain a clear scaling of $\langle N_2 \rangle (t)$, 
we proceed as follows.
First, we start a simulation from a mixture of independently distributed
empty sites and hosts with the equal density (i.e., 0.5 each).
After the system has approached a steady $S_{01}$ state,
we replace a randomly chosen empty site with a parasite
and continue the simulation until the stationary state is reached. 
Figure~\ref{additionalFig}(b) shows the time course of $\langle N_2 \rangle (t)$ 
for $\lambda_{12}=4$ and various values of $\lambda_{01}$, where the single parasite is added
at $t=0$.
Near the transition point, $\lambda_{01} \sim 0.591$, 
$\langle N_2 \rangle (t)$ seems to follow a power-law. 
The data for different values of $\lambda_{12}$
collapse onto a single curve with the DP critical exponents, separately 
for subthreshold and suprathreshold values of $\lambda_{01}$ (Fig.~\ref{additionalFig}(c)). 
Figure~\ref{additionalFig}(c)
suggests that the transition belongs to the DP universality class.

Note that $\langle N_2 \rangle (t)$ above the transition point saturates owing to a finite size effect. 
It is difficult to determine critical properties for large values of $\lambda_{12}$ 
because we would need increase $L$ to perform the dynamic scaling.
Nevertheless,
we believe that the $S_{01}$--$S_{012}$
transition belongs to the DP universality class even for larger $\lambda_{12}$. 

\section{summary}

We carried out numerical simulations for a three-state host-parasite
model on the square lattice.  The obtained phase diagram is shown in
Fig.~\ref{phasediagram}.  
Our numerical results suggest that the $S_{0}$--$S_{01}$ boundary
and the $S_{01}$--$S_{012}$ boundary are of the DP universality
class.
The
parasite-driven extinction occurs for large $\lambda_{01}$ and large
$\lambda_{12}$ in relatively small systems.
However, for a sufficiently large system, the three states coexist 
in the parameter region where the parasite-driven extinction occurs for a small system.
Therefore, the parasite-driven extinction is a finite size effect.
This prediction is consistent with the phase
diagram obtained from the PA  
but not with phase diagrams obtained from the mean field approximation and the i-PA.

\begin{acknowledgments}
We thank Alexei Tretiakov for valuable discussions.
N.K. acknowledges the support provided by the Japan Society for the Promotion
of Science through Grant-in-Aid for Scientific Research (C) (Grant No. 21540118).
N.M. acknowledges the support provided by MEXT, Japan 
through Grants-in-Aid for Scientific Research (Nos. 20760258 and 20540382). 
\end{acknowledgments}

%\begin{thebibliography}{99} 
%\end{thebibliography}

%\bibliography{ref3model}

\end{document}